\pdfoutput=1
\documentclass[
    aps,pre,twocolumn,
    reprint,
    superscriptaddress,
    nofootinbib,
    floatfix,
    amssymb,
    longbibliography
]{revtex4-2}
\usepackage[english]{babel}
\usepackage{amsmath}
\usepackage{amsfonts}
\usepackage{amssymb}
\usepackage[utf8]{inputenc}
\usepackage{nicefrac}
\usepackage[normalem]{ulem}
\usepackage{xcolor}

\definecolor{linkColor}{rgb}{0,0.3,0.7}
\usepackage[colorlinks=true,
            allcolors=linkColor,
            pdfborder={0 0 0},
            pdfencoding = auto
            ]{hyperref}

\usepackage{graphicx} 
\usepackage{tabularx} 
\usepackage{dcolumn}  
\usepackage{bm}       
\usepackage{hyperref}
\usepackage[capitalise]{cleveref}

\usepackage[inline]{enumitem} 
\usepackage{multirow}

\renewcommand{\vec}[1]{\bm{#1}}

\begin{document}

\title{Dimensionality reduction in bulk--boundary reaction--diffusion systems}

\author{Tom Burkart}
\thanks{These authors contributed equally.}
\affiliation{Arnold Sommerfeld Center for Theoretical Physics and Center for NanoScience, Department of Physics, Ludwig-Maximilians-Universit\"at M\"unchen, Theresienstra\ss e 37, D-80333 M\"unchen, Germany}
\author{Benedikt J.\ M\"uller}
\thanks{These authors contributed equally.}
\affiliation{Arnold Sommerfeld Center for Theoretical Physics and Center for NanoScience, Department of Physics, Ludwig-Maximilians-Universit\"at M\"unchen, Theresienstra\ss e 37, D-80333 M\"unchen, Germany}
\author{Erwin Frey}
\email{frey@lmu.de}
\affiliation{Arnold Sommerfeld Center for Theoretical Physics and Center for NanoScience, Department of Physics, Ludwig-Maximilians-Universit\"at M\"unchen, Theresienstra\ss e 37, D-80333 M\"unchen, Germany}
\affiliation{Max Planck School Matter to Life, Hofgartenstraße 8, D-80539 M\"unchen, Germany}

\date{\today}

\begin{abstract}%
Intracellular protein patterns regulate many vital cellular functions, such as the processing of spatiotemporal information or the control of shape deformations.
To do so, pattern-forming systems can be sensitive to the cell geometry by means of coupling the protein dynamics on the cell membrane to  dynamics in the cytosol.
Recent studies demonstrated that modeling the cytosolic dynamics in terms of an averaged protein pool disregards possibly crucial aspects of the pattern formation, most importantly concentration gradients normal to the membrane.
At the same time, the coupling of two domains (surface and volume) with different dimensions renders many standard tools for the numerical analysis of self-organizing systems inefficient.
Here, we present a generic framework for projecting the cytosolic dynamics onto the lower-dimensional surface that respects the influence of cytosolic concentration gradients in static and evolving geometries.
This method uses \textit{a priori} physical information about the system to approximate the cytosolic dynamics by a small number of dominant characteristic concentration profiles (basis), akin to basis transformations of finite element methods.
As a proof of concept, we apply our framework to a toy model for volume-dependent interrupted coarsening, evaluate the accuracy of the results for various basis choices, and discuss the optimal basis choice for biologically relevant systems.
Our analysis presents an efficient yet accurate method for analysing pattern formation with surface--volume coupling in evolving geometries.
\end{abstract}
\maketitle

\section{Introduction}\label{sec:introduction}%
Living organisms ensure their viability by precisely orchestrating a wide array of cellular functions, ranging from subcellular information processing to morphogenesis.
At the heart of these functions lie out-of-equilibrium molecular systems: proteins whose spatio-temporal organization in cells is driven by chemical interactions with other proteins and transport across the cell, commonly referred to as reaction--diffusion systems.
From these two building blocks, complex information processing systems can emerge that allow the transmission of chemical signals within the cell or encode temporal and spatial cues~\cite{Wolpert1981, Desai.Kapral2009, Burkart.etal2022}.
Popular examples of such pattern-forming protein systems are the Min system of \textit{Escherichia coli} which exhibits pole-to-pole oscillations \textit{in vivo}~\cite{Raskin.Boer1999} and spiral waves or labyrinth patterns \textit{in vitro}~\cite{Ramm.etal2019};
localization of the budding site via Cdc42 in \textit{Saccharomyces cerevisiae}~\cite{Chiou.etal2016};
or polarity establishment by PAR proteins in \textit{Caenorhabditis elegans}~\cite{Motegi.Seydoux2013}.

\begin{figure}[!t]
    \centering
    \includegraphics[width=\columnwidth]{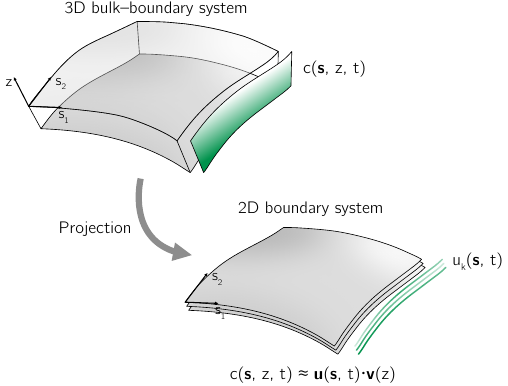}
    \caption{Schematic visualization of the projection of a time-dependent bulk field $c(\vec s, z, t)$ (indicated by green (gray) slice) in a three-dimensional bulk--boundary system onto the two-dimensional boundary of the domain via an approximation using multiple characteristic profiles $v_k(z)$ and corresponding time dependent membrane fields $u_k(\vec s, t)$ (indicated by green (gray) lines).}
    \label{fig:1}
\end{figure}

In dynamically deforming geometries, pattern-forming systems may gain the ability to engage in mechanochemical feedback loops that can give rise to even more interesting self-organizing dynamics and that play a key role in the regulation of many cellular functions.
Examples on the sub-cellular scale include the sensing and generation of membrane curvature by membrane-binding proteins~\cite{Walani.etal2014, Rogez.etal2019, Mahapatra.etal2021, Roux.etal2021, Wuerthner.etal2023} or the adaptive establishment of cell polarity~\cite{Hubatsch.etal2019}.
On the single-cell and tissue level, pattern formation in deforming geometries has been appreciated in the context of biochemical coordination of cell motility~\cite{Maree.etal2012, Camley.etal2017, Fu.etal2023}, cell shape changes~\cite{Miller.etal2018, Litschel.etal2018, Tamemoto.Noguchi2020, Hoerning.Shibata2019}, and the control of tissue morphogenesis~\cite{Brinkmann.etal2018, Mietke.etal2018}.
A central common feature of many pattern-forming systems in dynamic geometries is the coupling of two concurrently deforming domains -- the cell volume (or bulk) and the cell membrane (boundary) -- which models need to consider explicitly in order to accurately capture the mechanochemical coupling of the protein dynamics to the geometry~\cite{Burkart.etal2022}.

The realization of such systems in numerical simulations, however, offers two fundamental challenges: first, most numerical frameworks are not designed for solving reaction--diffusion dynamics in a deforming geometry, and including such deformations often abates the performance considerably.
Second, since the diffusion of membrane-bound proteins is typically slow compared to cytosolic proteins~\cite{Meacci.etal2006}, the length scale of protein patterns on membranes can be by orders of magnitude shorter than the length scale of protein gradients in the cell volume or the length scale of deformations.
This requires a high spatial resolution of the membrane and its vicinity, resulting in a fine mesh with high number of degrees of freedom that further slows down simulations.

A plethora of mathematical studies have focused on reaction--diffusion dynamics on the surface of a deforming geometry.
Popular methods that are used in such cases include specialized applications of the finite element method (FEM)~\cite{Dziuk.Elliott2007, Madzvamuse.etal2015, Chen.Ling2020}, level-set methods~\cite{Bergdorf.etal2010, Hansbo.etal2016}, and mesh-free approaches~\cite{Suchde.Kuhnert2019, Wendland.Kuenemund2020}.
However, it is often challenging to generalize these approaches to systems with bulk--boundary coupling, i.e., systems with dynamics both in the volume and on the surface of the deforming geometry, as this requires handling the deformation effects in two different domains simulatneously and in a self-consistent manner~\cite{Lervag.Lowengrub2015}.
Many studies of biological systems handle this limitation by averaging the volume dynamics either over infinitesimal surface patches (projection) or over the entire volume (reservoir)~\cite{Loose.etal2008, Bement.etal2015}.
While this approach yields acceptable results when the volume dynamics are purely diffusive~\cite{Paquin-Lefebvre.etal2020}, it was recently demonstrated that accounting for the bulk dynamics is crucial for accurate predictions in the presence of bulk reactions~\cite{Halatek.etal2018, Brauns.etal2021, Wuerthner.etal2022}.
This is because bulk reactions generically lead to protein concentration gradients in the cytosol which, by virtue of the bulk--boundary coupling, directly affect the protein dynamics on the membrane and thereby provide the basis for one type of geometry sensing~\cite{Burkart.etal2022}.
In studies committed to faithfully representing the reaction--diffusion dynamics in deforming geometries, the phase--field method has emerged as a promising strategy~\cite{Levine.Rappel2005, Li.etal2009, Teigen.etal2009, Yu.etal2012, Marth.Voigt2014, Raetz.Roeger2014, Abels.etal2015, Camley.etal2017, Moure.Gomez2020, Valizadeh.Rabczuk2019}.
Phase--field models represent the dynamic geometry as an indicator function, allowing for arbitrary shape changes and even topological changes~\cite{Biben.etal2005, Li.etal2009}.
However, this benefit comes at the cost of either requiring fine mesh resolution everywhere in the simulated domain or using adaptive mesh refinement~\cite{Raetz.Voigt2006, Elliott.etal2011}.
Furthermore, coupling bulk dynamics to surface dynamics in phase--field models raises additional challenges that are subject of ongoing development~\cite{Poulsen.Voorhees2018, Yu.etal2020}.

Here, we present an approach that exploits \textit{a priori} knowledge about cytosolic gradients for modeling reaction--diffusion dynamics with bulk--boundary coupling in deforming geometries.
Rather than solving the ensuing partial differential equations (PDEs) in a meshed bulk, we propose to project the bulk dynamics onto the surface of the deforming geometry by decomposing the bulk dynamics into a few dominant basis functions, essentially reducing the spatial dimension of the system by one [Fig.~\ref{fig:1}].
We show how to design projection methods based on information about the steady state of the system at hand and compare three different techniques -- based on step functions, polynomials, and exponential functions -- with respect to their accuracy in flat and in deforming geometries.
Our analysis demonstrates that good approximations of the actual dynamics can be achieved already with a single nonlinear basis function, whereas a naive averaging of the bulk dynamics often fails to capture the relevant dynamics entirely.
Furthermore, we find that the projection method outperforms standard FEM approaches by an up to five-fold speedup in computation time, with most significant improvements in systems with large volume.

This paper is structured as follows:
In Sec.~\ref{sec:model}, we first introduce a generic projection method in a static one-dimensional geometry and subsequently generalize our approach to dynamic higher-dimensional geometries.
We show how to transfer techniques from FEM implementation to arbitrary basis choices and how a PDE defined on a volume domain can be approximated by a set of PDEs defined only on the volume's boundary.
In Sec.~\ref{sec:applications} we test our approach on a toy model that shows geometry-dependent arrested coarsening and compare different projection methods based on the quality of the approximation and the computational efficiency.
We conclude with a concise summary and an outlook.

\section{Bulk projection}\label{sec:model}
In this section, we introduce a method to project the dynamics of proteins in the bulk onto the membrane for a flat geometry.
We illustrate this method using a generic model for a single protein species that can bind to and unbind from the membrane and are degraded linearly in the bulk.
We then extend the projection method to dynamically deforming membranes.

\subsection{Static geometry}%
\begin{figure}[!tb]
    \centering
    \includegraphics{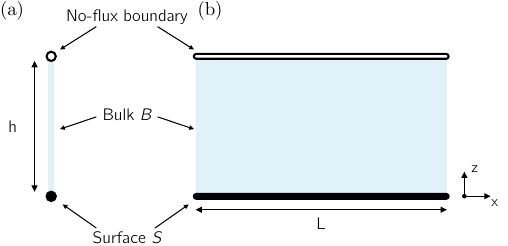}
    \caption{Schematic visualization of flat geometries.
    (a)~A one-dimensional bulk domain $\mathcal{B}$ (cytosol, blue/gray) of size $h$ is bounded by a reactive surface $\mathcal{S}$ (membrane, filled circle) on one side and by a no-flux boundary on the other side (open circle).
    (b)~For a two-dimensional bulk domain of height $h$ and width $L$, the surface $\mathcal{S}$ is a one-dimensional line.}
    \label{fig:flat_geometry_sketch}
\end{figure}
Consider a one-dimensional bulk $\mathcal{B}$ of height $h$ [Fig.~\ref{fig:flat_geometry_sketch}a]. 
The bulk (cytosolic) protein concentration on this line, denoted by $c(z, t)$, is assumed to obey no-flux (Neumann) boundary conditions at the top of the line ($z=h$) and Robin boundary conditions -- where reactive fluxes are balanced by diffusive fluxes onto the boundary -- at the bottom of the line (membrane at $z=0$, denoted by $\mathcal{S}$).
The protein concentration on the membrane is denoted by $m(t)$.

A generic (non--mass-conserving) bulk--boundary reaction--diffusion system for a single protein species with linear degradation in the bulk can be written as
\begin{subequations}
\label{eq:II.1}
\begin{align}
    \partial_t c(z,t) &= D_c \, \partial_z^2 c(z,t) - \lambda \cdot c(z,t) \, , \\
    \partial_t m(t) &= f_0(m(t), c(0,t)) \, , 
\end{align}
\end{subequations}
where the bulk dynamics are coupled to the membrane via Robin and no-flux boundary conditions
\begin{subequations}
\label{eq:II.2}
\begin{align}
    - D_c \, \left. \partial_z c(z,t) \right|_{z=0} &= - f_0(m(t), c(0,t)) \, , \\
    D_c \, \left. \partial_z c(z,t) \right|_{z=h} &= 0 \, . 
\end{align}    
\end{subequations}
Here, $D_c$ denotes the diffusion constant in the bulk, $\lambda$ is a cytosolic degradation rate, and $f_0(m(t), c(0,t))$ denotes the reactive fluxes at the surface $\mathcal{S}$.
A generic choice for the reaction at the boundary is
$$f_0(m(t), c(0,t)) = a(m(t)) \cdot c(0,t) - d(m(t)) \cdot m(t)  \,,$$
where the attachment $a$ and detachment $d$ to and from the surface $\mathcal S$ may include (autocatalytic) nonlinear interactions.

To project the dynamics in a one-dimensional bulk onto the zero-dimensional surface $\mathcal{S}$, we aim to rewrite the partial differential equation (PDE) for the bulk dynamics in Eq.~\eqref{eq:II.1} as a set of $N$ conveniently chosen ordinary differential equations (ODEs) that approximate the exact solution.
Analogous to reaction--diffusion equations that can be split into reaction and diffusion parts~[Eq.~\eqref{eq:II.1}] and boundary conditions [Eq.~\eqref{eq:II.2}], we thus aim for a set of ODEs in the form of
\begin{equation}
    \partial_t u_k(t) = g_\text{React} + g_\text{Diff} + g_\text{BC}
\end{equation}
for $k \in \{0,\ldots,N{-}1\}$.
The symbolic functions $g = g\bigl(\{u_k(t)\}, m(t)\bigr)$ are placeholders for the contributions to the ODE stemming from the bulk reactions, bulk diffusion, and the boundary conditions, respectively.
How can one derive such a set of ODEs (or, in higher dimensions, PDEs)?

To answer this question, we draw inspiration from numerical methods for solving PDEs, specifically the \emph{finite element method} (FEM).
The general idea of FEM approaches is to approximate the exact dynamics by a linear combination of contributions derived for simple basis functions (finite elements), where each basis function represents the field value at a specific point in the meshed simulation domain~\cite{Zienkiewicz.etal2005, Brenner.Scott2008}.
Since, for generic applications, the dynamics of the concentration fields are not known \textit{a priori}, it is vital for an efficient FEM solver to use simple basis functions with little overlap on a well-resolved mesh.
For physical systems, however, certain properties of the field dynamics can be derived \textit{a priori}.
For example, in a reaction--diffusion system with linear bulk reactions as in Eq.~\eqref{eq:II.1}, the characteristic diffusive length scale in the bulk is $\ell = \sqrt{D_c/\lambda}$~\cite{Frey.Brauns2022}.
In the following, we show how a more sophisticated choice of basis functions that exploits such knowledge can simplify the problem and reduce the complexity of numerical implementations solving the system's dynamics.

\begin{figure*}
    \centering
    \includegraphics[width=\textwidth]{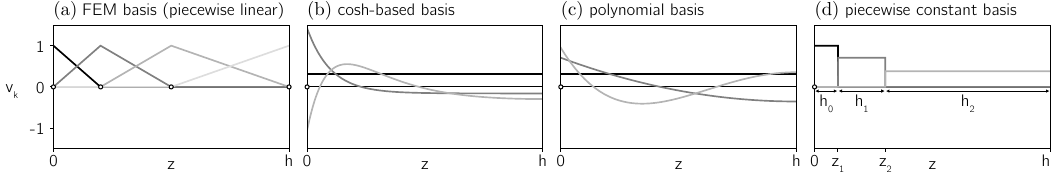}
    \caption{Comparison of different basis choices. Individual basis functions are indicated by different shades of gray.
    (a)~Typical choice of piecewise linear basis functions used in FEM with irregular mesh (mesh points indicated by open circles).
    (b)~Orthonormalized basis derived from $v_k = \cosh\bigl((h{-}z)/\ell_k\bigr)$ with $\ell_k = \{\infty, 1, 2 \}$. This basis choice can capture strong gradients close to the membrane ($z=0$) paired with weak gradients at large heights, but not vice versa.
    (c)~Orthonormalized basis derivd from $v_k = (z-h)^{2k}$.
    (d)~Orthonormalized piecewise constant basis with increasing layer heights $h_k$ for fine resolution of the membrane-proximal region.}
    \label{fig:II_basis}    
\end{figure*}

While FEM in the context of numerical simulations typically uses a single basis function per mesh point in the simulated domain~\cite{Brenner.Scott2008}, we pursue a different approach using multiple basis functions $v_k(z)$ anchored to the boundary of the domain [Fig.~\ref{fig:II_basis}].
The different basis functions are designed to reflect the geometry of the bulk domain and to ensure that the relevant aspects of concentration gradients perpendicular to the boundary can be captured appropriately.
In contrast to FEM, the coefficients $u_k(t)$ are defined for mesh points on the membrane so that all bulk variations are captured by the height-dependent basis functions $v_k(z)$.
With this (incomplete) basis, the actual bulk dynamics are approximated by
\begin{equation}
    c(z,t) \approx \sum\limits_k u_k(t) \cdot v_k(z) \, .
\end{equation}
Inserting this ansatz into the reaction--diffusion equations~\eqref{eq:II.1} and~\eqref{eq:II.2} yields, after partially integrating the diffusion term and taking the inner product $\langle \cdot, \cdot  \rangle$ with a basis function $v_k$ (Galerkin projection~\cite{Brenner.Scott2008}), a set of coupled ODEs for the coefficients $u_k$ (detailed derivation in Appendix~\ref{app:derivation}):
\begin{align}
    G_{kl} \, \partial_t u_k(t) &= \underbrace{- f_0\bigl(m(t), \{u_k(t)\} \bigr) \, v_l(0) }_{g_\text{BC}}  + \nonumber \\
    & \qquad \underbrace{-D_c \, A_{ml} \, u_m(t)}_{g_\text{Diff}} -  \underbrace{\lambda\, G_{lm} \, u_m(t)}_{g_\text{React}} \, . \label{eq:II.9}
\end{align}
Here and in the following, we use Einstein sum convention over double indices.
With the inner product
$$
    \langle v(z), w(z) \rangle = \int_0^h\!\!\!\mathrm d z \; v(z) \, w(z)
$$
the \emph{mass matrix} (or Gram matrix) $G_{kl}$ and \emph{stiffness matrix} $A_{kl}$ are defined as
\begin{align}
    G_{kl} &= \langle v_k(z) , v_l(z) \rangle \, , \nonumber \\
    A_{kl} &= \langle \partial_z v_k(z) , \partial_z  v_l(z) \rangle \, . \label{eq:II.stiffness_matrix}
\end{align}
From here on we will omit all function arguments unless needed for clarity.

The set of ODEs in Eq.~\eqref{eq:II.9}, together with the basis $\{v_k\}$, yield an approximation of the time evolution of the bulk field $c$, where the quality of the approximation critically depends on the basis choice.
What is a good choice for the basis $\{v_k\}$?
In general, the basis needs to be chosen such that it can capture the main features of the bulk dynamics.
Specifically, this requires that both the field $c$ as well as its gradient $\partial_z c$ are approximated well by the projection $u_k \, v_k$.
Indeed, it has been shown that the coefficients $u_k$ obtained from $u_k = (G^{-1})_{kl} \langle v_l, c \rangle$ yield the best possible approximation for the field $c$ for a given basis choice $\{v_k\}$~\cite{Brenner.Scott2008}.
However, no such statement exists for the gradient $\partial_z c$, and the same coefficients $u_k$ can result in a low-quality approximation of the gradient for improper basis choices.
Here we present two ways to resolve this problem:
\begin{enumerate*}[label=(\roman*)]
    \item by choosing a basis that is expected \textit{a priori} to yield good approximations for both field $c$ and gradient $\partial_z c$, or
    \item by making additional choices for the gradients and thereby correct for the low-quality approximation of the gradients.
\end{enumerate*}

The former approach strongly depends on the physical problem that is being studied.
For example, for protein reaction--diffusion systems with linear bulk reactions as specified in Eq.~\eqref{eq:II.1} and no-flux boundary conditions at $z{=}h$, the steady state distribution can be derived as $c(z) \sim \cosh\bigl((h-z)/\ell\bigr)$, with $\ell = \sqrt{D_c/\lambda}$~\cite{Frey.Brauns2022}.
In this case, a natural choice for the basis would be a set of hyperbolic cosine functions on varying length scales [Fig.~\ref{fig:II_basis}b], e.g.,
$$\{v_k(z)\} = \left\{1,\, \cosh\left(\frac{h-z}{\ell} \right), \,\cosh\left(\frac{h-z}{2\ell} \right) , \, \ldots \right\} \,.$$
A mathematically more tractable choice for the same system is $\{v_k(z)\} = \{(h{-}z)^k\}$, which corresponds to a polynomial fit to the field $c$ and can be expected to approximate shallow gradients well [Fig.~\ref{fig:II_basis}c].

For the latter approach revolving around making additional choices for the gradients, we only discuss a basis composed of orthogonal step functions, as this choice will prove highly useful for deforming geometries later on.
For this, the bulk is divided in $N$ sections of height $h_k$, where each basis function $v_k$ takes the form of a rectangle function [Fig.~\ref{fig:II_basis}d].
Since the derivative of these basis functions at the interfaces $z_k$ are not well-defined, we instead approximate the gradients in between the centers of the step functions as an additional set of piecewise constant functions, as derived in Appendix~\ref{app:step}.
To highlight this additional approximation, we indicate auxiliary stiffness matrices constructed from such an artificial gradient choice by a bar over the symbol ($\bar A_{kl}$).
For piecewise constant gradients, the resulting set of ODEs for the coefficients $u_k$ then reads [Appendix~\ref{app:step}]
\begin{subequations}
\begin{align}
    \partial_t u_k &= - \delta_{k0} \, f_0 \, v_0(0)  - D_c \,   \bar A_{kl} \, u_l + \lambda \, u_k  \,, \label{eq:II.10} \\
    \bar A_{kk} &= \frac{2}{h_k + h_{k{+}1}}+ \frac{2}{h_k + h_{k{-}1}} \,,  \\
    \bar A_{k,k{\pm}1} &= -\frac{2}{h_k + h_{k{\pm}1}} \,,    
\end{align}     
\end{subequations}
where $\delta_{kl}$ is the unit matrix and all other entries $\bar A_{kl} =0$.
It is instructive to compare this choice with standard finite element methods on a predefined mesh: in FEM implementations, basis functions are typically non-zero only in the direct vicinity of a specific mesh point, similar to the step functions that are non-zero only for a fraction of the bulk domain~[Fig.~\ref{fig:II_basis}d]~\cite{Brenner.Scott2008}.
Furthermore, FEM basis functions are typically piecewise linear, such that their gradients are piecewise constant and thus the stiffness matrix takes a similar form as in Eq.~\ref{eq:II.10}.
Rather than fully committing to the FEM approach, we here stick to piecewise constant basis functions and the auxiliary stiffness matrix $\bar A_{kl}$ since this will greatly simplify the generalization to deforming geometries in the following section.

\subsection{Dynamically deforming geometries}\label{sec:method:deforming}
\begin{figure}
    \centering
    \includegraphics[width=\columnwidth]{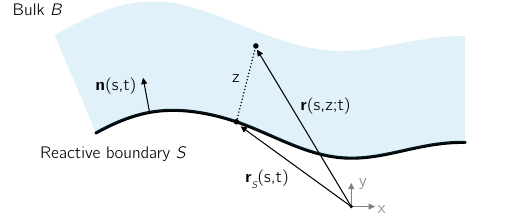}
    \caption{Schematic visualization of a deforming geometry.
    The bulk domain $\mathcal B$ (cytosol, blue/gray) is parametrized by a vector $\vec r(s,z; t)$, with $z$ the distance from the reactive boundary $\mathcal S$ (membrane, thick black line) measured along the normal vector $\vec n (s,t)$.
    The boundary itself is parametrized by $\vec r_\mathcal{S}(s,t) = \vec r(s,0; t)$.}
    \label{fig:geometry}
\end{figure}
We now generalize the idea of bulk projection to dynamically deforming geometries.
Consider a two-dimensional domain $\mathcal{B}$ parametrized by  $\vec r(s,z)$, where $s$ denotes the (arc-length) parametrization of the one-dimensional membrane $\mathcal{S}$, $\vec r_\mathcal{S} (s,t)$, and $z \in [0,h]$ is the distance from the membrane [Fig.~\ref{fig:geometry}]:
\begin{equation}\label{eq:II.parametrization}
    \vec r(s,z; t) = \vec r_\mathcal{S} (s,t) + z \, \vec{\hat n}(s,t) \, .
\end{equation}
The vector $\vec{\hat n}$ is the normal vector on the membrane.
The bulk dynamics specified in Eq.~\eqref{eq:II.1} in such a dynamic geometry read~\cite{Wuerthner.etal2023}
\begin{subequations}
\begin{align}
    \frac{1}{\sqrt{g}} \partial_t \bigl(\sqrt{g} \, c\bigr) &= D_c \, \Delta_\text{LB} c + \lambda \cdot c \, , \\
    \Delta_\text{LB} c &= \frac{1}{\sqrt{g}} \partial_i \bigl( \sqrt{g} g^{ij} \partial_j c \bigr) \, , 
\end{align}    
\end{subequations}
with the Laplace-Beltrami operator $\Delta_\text{LB}$, the metric tensor $g_{ij}$, and the square root of the metric determinant $\sqrt{g} \equiv \sqrt{\det(g)}$.
Furthermore, we generalize the inner product, the Gram matrix $G$, and the stiffness matrix $A$ to the deformable geometry:
\begin{align}
    \langle u(z), v(z) \rangle^{ij} &= \int_0^h \!\!\! \mathrm dz \; \sqrt{g(s,z;t)}\, g^{ij}(s,z;t) \, u(z) \cdot v(z) \, , \nonumber \\
    G_{kl}^{ij}(s,t) &= \langle \, v_k, v_l \rangle^{ij} \, , \label{eq:II:deformable_projection}\\
    A_{kl}^{ij}(s,t) &= \langle \partial_z v_k, \partial_z v \rangle^{ij}  \, , \nonumber
\end{align}
so that $G_{kl} \equiv G^{zz}_{kl} = \langle v_k, v_l \rangle^{zz}$, and analogous for $A_{kl}$.
In Appendix~\ref{app:derivation}, we use these definitions to derive the weak formulation and Galerkin projection~\cite{Brenner.Scott2008} of the bulk dynamics in a deforming geometry as specified in Eq.~\eqref{eq:II.parametrization}.
The resulting generalized set of PDEs for the coefficients reads
\begin{align}
    G_{kl} \partial_t u_k &= \underbrace{-u_m \, (\partial_t G)_{lm}}_{g_\text{Geom}} + \underbrace{D_c \, \partial_s \bigl(G^{ss}_{lm} \, \partial_s u_m \bigr)}_{g_\mathcal{S}} + \label{eq:II.projection_time_dependent}  \\
    &\quad + D_c \bigl[\sqrt{g} \partial_z c \, v_l \bigr]_0^h - D_c \, u_m \, A_{lm} + \lambda \, u_m \, G_{lm} \, .\nonumber
\end{align}
In comparison to the set of ODEs in Eq.~\eqref{eq:II.9}, two new terms have emerged in addition to the redefined coupling matrices:
The contribution $g_\text{Geom}$ accounts for deformations of the domain (dilation and compression), and $g_\mathcal{S}$ accounts for the diffusion parallel to the membrane.

Strikingly, this formulation holds for any basis where all elements $\{v_k\}$ are weakly differentiable on $[0,h]$.
In addition, the formulation can be adjusted for a basis of orthogonal step functions by appropriate rescaling of the stiffness matrix $\bar A_{kl}$ [Appendix~\ref{app:step}].
This offers significant advantages compared to an explicit simulation of the bulk dynamics:
first, the number of degrees of freedom can be reduced significantly by avoiding a meshed bulk domain.
Second, similar to other methods that use an explicit parametrization of the geometry, our approach also makes a dynamic remeshing of the deforming bulk obsolete.
These advantages come at the cost of having to settle on a basis \textit{a priori}, where the quality of the approximation critically depends on the basis choice.

With the projection step in Eq.~\eqref{eq:II.projection_time_dependent}, one therefore converts a reaction--diffusion system with coupled PDEs in the bulk and on the membrane into a set of PDEs that are defined only on the membrane.
This approach yields an effective reduction of the system's spatial dimension (e.g., from a three-dimensional bulk to a two-dimensional membrane) while accounting for bulk gradients in an approximate manner and thereby respects the effect of the bulk geometry on the protein dynamics.
The accuracy of this approximation depends on the basis choice $\{v_k\}$.
These benefits come at the cost of introducing one additional PDE on the membrane for each basis function.

\section{Examples}\label{sec:applications}
\begin{figure}
    \centering
    \includegraphics[width=\columnwidth]{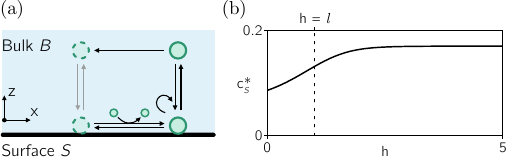}
    \caption{(a)~Schematic reactions in the geometry-sensitive coarsening toy model. A protein can bind to an unbind from the membrane in its active state (solid). In the bulk, the protein undergoes deactivation (dashed). On the membrane, inactive proteins can be activated assisted by active bulk proteins. The inactive proteins are assumed to be abundant (reservoir) and undergo linear binding to and unbinding from the membrane, thereby maintaining a constant concentration on the membrane.
    (b)~The steady-state concentration of the bulk field close to the membrane $c^*_\mathcal{S}$ increases with bulk height for  $h \lesssim \ell $ but is approximately constant for $h \gg l$.}
    \label{fig:toy_model}
\end{figure}
In the following, we present and discuss an explicit application of our dimensionality reduction method for a simple toy model, with particular focus on comparing different basis choices and on the influence of the bulk geometry.
For this, we extend a previously studied model showing interrupted coarsening by introducing an explicit sensitivity to the bulk geometry~\cite{Brauns.etal2020, Brauns.etal2021a}.
This choice allows to easily quantify the effect of the bulk geometry and, most importantly, how well the bulk dynamics are captured by the projection method by means of the pattern length scale after coarsening is interrupted.
In this model, we consider a protein that switches between an active and an inactive state and prevails either in the bulk or on the membrane [Fig.~\ref{fig:toy_model}a].
In the active state, the protein can bind and recruit itself to the membrane, and undergoes enzyme-mediated detachment.
Active proteins deactivate both on the membrane and in the bulk, but are assumed to be reactivated only on the membrane and assisted by membrane-proximal active bulk proteins.
For simplicity, we furthermore assume that the inactive proteins comprise an abundant reservoir maintaining a constant concentration on the membrane and are therefore not modeled explicitly.
The reaction--diffusion equations for the active proteins on the membrane $m(x,t)$ and in the bulk $c(\vec x, t)$ then read~\cite{Brauns.etal2020, Brauns.etal2021a}
\begin{subequations}
\label{eq:III:toy_model} 
\begin{align}
    \partial_t m &= D_m \, \partial_x^2 m + f(m, \left.c\right|_\mathcal{S}) + \zeta (m, \left.c\right|_\mathcal{S}) \, , \\
    \partial_t c &= D_c \, \nabla^2 c - \lambda \, c \,, 
\end{align}    
\end{subequations}
with Robin boundary conditions
\begin{equation}
    D_c \, \left. \vec n \cdot \nabla c \right|_\mathcal{S} = - f(m, \left.c\right|_\mathcal{S})  \, ,
\end{equation}
where the mass-conserving binding kinetics of active proteins are given by
\begin{equation}
    f(m, \left.c\right|_\mathcal{S}) = (1+m) \cdot \left.c\right|_\mathcal{S} - \frac{m}{1+m} \, .
\end{equation}
Here, $ \left.c\right|_\mathcal{S} = c(x, z{=}0)$ denotes the cytosolic concentration at the membrane, and we have set all reaction rates and parameters to $1$ for simplicity.
In addition, the non-mass-conserving part corresponding to the (de-)activation of proteins on the membrane (by exchange with an abundant reservoir [Fig.~\ref{fig:toy_model}a]) is
\begin{equation}
    \zeta(m,\left.c\right|_\mathcal{S}) = p \, \left.c\right|_\mathcal{S} - \epsilon \, m 
\end{equation}
with an effective activation rate $p$ and deactivation rate $\epsilon$.
With Robin boundary conditions at the membrane and linear degradation, the bulk concentration has an associated length scale $\ell = \sqrt{D_c/\lambda}$ characterizing gradients at the steady state~\cite{Frey.Brauns2022}.
For simplicity, we further nondimensionalize this toy model by enforcing $\ell = 1$, so that the system height $h$ as well as all other distances will always be given in units of the characteristic length scale (full parameter list in Table~\ref{tab:params}).

\begin{figure}
    \centering
    \includegraphics[width=\columnwidth]{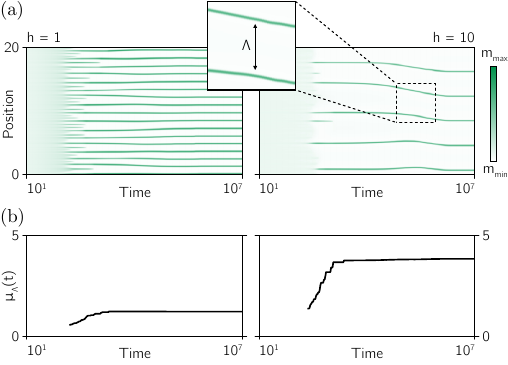}
    \caption{(a)~Typical kymographs $m(x,t)$ of the model specified in Eqs.~\eqref{eq:III:toy_model} for $h=1$ (left column) and for $h=10$ (right column). The system undergoes interrupted coarsening and subsequent wavelength selection. Time is shown on a log scale. The peak distance $\Lambda$ is indicated in the inset.
    (b)~Mean peak distance $\mu_\Lambda$ for the two kymographs shown in panel~(a).
    At large system heights, coarsening is interrupted at considerably higher mean peak distances than for small system heights.}
    \label{fig:kymograph}
\end{figure}

In Fig.~\ref{fig:kymograph}a, we show kymographs of two realizations of this system on a domain of width $L=20$ and at bulk heights $h=1$ and $h=10$ obtained from FEM simulations that explicitly account for the bulk dynamics.
The system undergoes interrupted coarsening, where initially multiple peaks of high protein concentration on the membrane (green) form that subsequently merge or vanish (\emph{coarsening}) until only a few peaks remain (\emph{interruption}).
After coarsening is interrupted, peaks rearrange to accomodate approximately equidistant spacing $\Lambda$ between them.
A signature of interrupted coarsening is the mean peak distance $\mu_\Lambda$ saturating at finite values [Fig.~\ref{fig:kymograph}b].
For the reaction--diffusion system in Eq.~\eqref{eq:III:toy_model}, the coarsening process is predominantly controlled by the (de-)activation of proteins on the membrane $\zeta(m, \left. c\right|_\mathcal{S})$~\cite{Brauns.etal2021a} and thus depends on the membrane-proximal bulk concentration $\left. c \right|_\mathcal{S}$.
Since this membrane-proximal bulk concentration in turn depends on the system height $h$ [Fig.~\ref{fig:toy_model}b], the coarsening interruption is sensitive to the bulk geometry.
Importantly, similar to the steady-state concentration $c^*_\mathcal{S}(h)$ shown in Fig.~\ref{fig:toy_model}b, we expect a nonlinear dependence of the mean peak distance on the system height $\mu_\Lambda(h)$, with an approximately linear dependence for $h \lesssim \ell$ and no variation for $h \gg \ell$.

\begin{figure*}
    \centering
    \includegraphics[width=\textwidth]{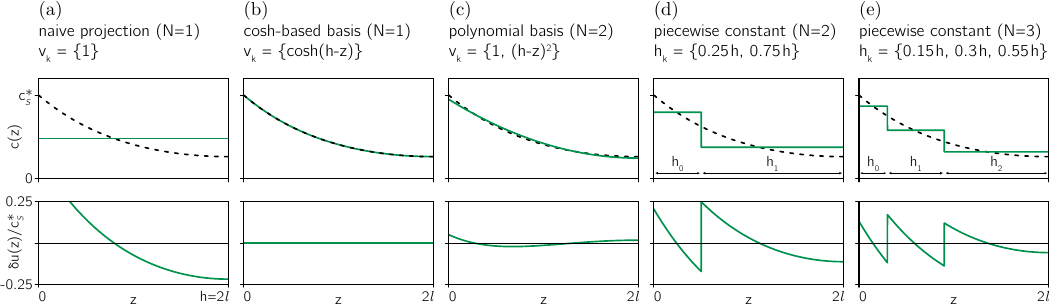}
    \caption{Approximation of the steady-state bulk concentration profile $c(z)$ for the system specified in Eq.~\eqref{eq:III:toy_model} (black, dashed) using finite bases $v_k(z)$ for bulk height $h=2\ell$. Top row shows the best approximation by least squares $u(z) = u_k \, v_k(z)$ (green/gray), bottom row shows the deviation from the exact profile $\delta u (z)= c(z) - u(z)$.
    (a)~Naive projection averaging over the bulk height ($N=1$).
    (b)~A basis using a hyperbolic cosine fully captures the steady-state concentration profile ($N=1$).
    (c)~A polynomial basis yields good approximations for sufficiently small bulk heights ($N=2$).
    (d,e)~A piecewise constant basis yields considerable deviations from the exact profile which can be reduced by increasing the number of basis functions/``layers'' ($N\geq 2$).
    }
    \label{fig:basis_choices}
\end{figure*}

\subsection{Basis choices}
As a naive direct projection of the bulk dynamics onto the membrane one may choose to average the bulk field $c$ over the $z$-direction.
This corresponds to choosing a basis $v_k(z) = \{1 \}$ with a constant function as the only basis element [Fig.~\ref{fig:basis_choices}a].
By definition, this ansatz disregards all gradients in the bulk field.
These gradients are shallow for small bulk heights $h < \ell$, however they become significant when the bulk height is larger than the characteristic length scale of bulk gradients, $h > \ell$.
Consequently, this naive projection captures the average pattern length scale $\mu_\Lambda$ only for small bulk heights but fails for large bulk heights [Fig.~\ref{fig:compare_basis_flat}a].

In the next step, we extend the projection basis by one additional element to capture the bulk gradients, i.e., variations in $z$-direction.
To specify this additional basis element, we use the system's laterally homogeneous steady state (constant parallel to the membrane), which can be calculated analytically for the linear reactions as defined in Eq.~\eqref{eq:III:toy_model} \cite{Frey.Brauns2022}:
\begin{equation}
    c^*(\vec x) = c^*(z{=}0) \cdot  \frac{\cosh \bigl( \frac{h-z}{\ell} \bigr) }{\cosh(h/\ell)} \, ,
\end{equation}
where $c^*(z{=}0) = c^*_\mathcal{S}$ is the steady state concentration at the membrane.
In the rescaled system with $\ell=1$, the steady state profile can be represented exactly by a one-dimensional basis $v_k(z) = \{\cosh(h{-}z)\}$.
Using this basis, gradients on the length scale of the characteristic scale $\ell$ are captured exactly [Fig.~\ref{fig:basis_choices}b].

In between these two limiting cases (naive averaging using $v_k=\{1\}$ and full recovery of the steady state using $v_k=\{\cosh(h{-}z)\}$) a plethora of other basis choices can be imagined.
The suitability of a specific basis choice depends on the desired application.
For example, while a hyperbolic cosine-derived basis captures the steady state bulk profile exactly in a flat geometry, no closed-form expressions for required Gram tensor $G^{ij}_{kl}$ and stiffness tensor $A^{ij}_{kl}$ as defined in Eq.~\eqref{eq:II:deformable_projection} may exist for deformed geometries with non-trivial metric $g_{ij}$, making them unsuitable for numerical implementations.
In particular, it is desirable to calculate the Gram and stiffness tensors as functions of time -- which requires the inner products to be solved analytically -- since this allows a more flexible implementation compared to recalculating the tensors after each time step by solving the inner products numerically.
To achieve this, one may expand the exact profile as a power series that respects the no-flux boundary condition at $z{=}h$, i.e., using a polynomial basis with even powers [Fig.~\ref{fig:basis_choices}c].

Alternatively, one may choose a set of piecewise constant basis functions that splits the bulk into $N$ distinct sections of height $h_k$ to approximate the bulk concentration profiles [Fig.~\ref{fig:basis_choices}d].
Similar to finite element methods, the coefficients $u_k$ then denote the average value of the bulk field in the $k^\text{th}$ section.
By tuning the section heights $h_k$ the approximation can be refined in regions with comparably sharper bulk gradients.
For example, the membrane-proximal region for the toy model in Eq.~\eqref{eq:III:toy_model} requires a finer resolution along the $z$-direction than regions at the opposing no-flux boundary due to the shallow gradient at $z{=}h$ [Fig.~\ref{fig:basis_choices}d].
The accuracy of the approximation can further be improved by increasing the number of basis functions [Fig.~\ref{fig:basis_choices}e].

The comparison of the approximated steady-state concentration profile with the exact counterpart can provide preliminary intuition about appropriate basis choices.
However, when studying the dynamic case, additional aspects of the bulk concentration may become relevant.
For example, the hyperbolic cosine in Fig.~\ref{fig:basis_choices}b can capture the long-term (steady state) dynamics but will not account for sharp bulk gradients as the steady state is approached.
How do the basis choices perform for approximating the protein \emph{dynamics}?

\subsection{Flat and static geometry}
To address this question, we first focus on flat and static geometries [Fig.~\ref{fig:flat_geometry_sketch}].
Since the projection of the bulk dynamics onto the membrane is a qualitative approximation, it is not expected that patterns obtained from the full system and from the approximation match quantitatively.
Instead, we are interested in capturing the qualitative characteristics of the patterns and compare the pattern's statistical properties to measure the quality of a basis choice.
For the interrupted coarsening system in Eq.~\eqref{eq:III:toy_model}, the most important statistical property is the mean value $\mu_\Lambda$ of the distance between peaks $\Lambda$ after coarsening is interrupted [Fig.~\ref{fig:kymograph}].

\begin{figure}[!t]
    \centering
    \includegraphics[width=\columnwidth]{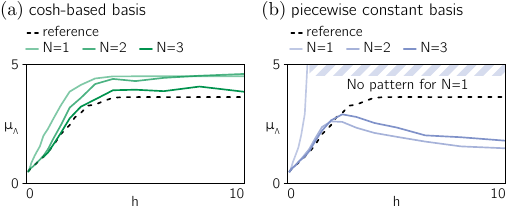}
    \caption{Comparison of the mean peak distance $\mu_\Lambda$ after coarsening using the projection method with different basis choices. Reference values obtained from explicit FEM simulations including the bulk are shown as dashed lines.
    (a)~Hyperbolic cosine-derived basis with increasing number of basis functions indicated by darker shades of green (dark gray).
    (b)~Piecewise constant basis with increasing number of basis functions indicated by darker shades of blue (light gray). The case $N=1$ corresponds to naive averaging of the bulk dynamics and predicts no pattern for $h > 1$, indicated by the striped region.
    }
    \label{fig:compare_basis_flat}
\end{figure}

For this, we solved the PDEs for the full system and for the projected counterparts on a domain of length $L=50$ with varying bulk height $10^{-1} \leq h \leq 10^1$ until coarsening was interrupted (total simulation time $T=10^7$; random perturbations around the homogeneous steady state as initial conditions) and extracted the mean peak distance $\mu_\Lambda$ at the final time point.
To reduce artifacts from initial conditions, the results are averaged over five repeats with varying initial conditions each.
In Fig.~\ref{fig:compare_basis_flat}a, we compare these results for the full system (\emph{reference}) as well as three different basis choices derived from hyperbolic cosines:
\begin{align*}
    N&=1: & v_k &=\bigl\{\cosh(h{-}z)\bigr\} \,,\\
    N&=2: & v_k &=\bigl\{1, \cosh(h{-}z)\bigr\} \,,\\
    N&=3: & v_k &=\bigl\{1, \cosh(h{-}z), \cosh\bigl(\tfrac{h{-}z}{\ell_2}\bigr)\bigr\} 
\end{align*}
with a variable length scale $\ell_2$ in the latter case.
All basis choices reliably capture the characteristic shape of the reference $\mu_\Lambda(h)$, however they consistently overestimate the true mean peak distance [Fig.~\ref{fig:compare_basis_flat}a].
Importantly, even though the basis with $v_k^{(N{=}1)} = \{\cosh(h{-}z)\}$ can capture the steady state exactly as shown in Fig.~\ref{fig:basis_choices}b, additional basis functions are required to reduce the deviation from the reference in the dynamic case.

Figure~\ref{fig:compare_basis_flat}b shows the quantification of mean peak distances using a piecewise constant basis for projecting the bulk dynamics [Fig.~\ref{fig:basis_choices}a,d,e].
The case $N=1$ corresponds to a naive averaging of the bulk dynamics and deviates significantly from the reference value $\mu_\Lambda(h)$ already for small bulk heights $h< \ell$.
For $h\geq \ell$, the naive projection predicts no patterns at all (shaded region) and is insufficient for capturing the bulkd--boundary coupled reaction--diffusion dynamics.
The predictions can again be improved by introducing additional basis functions, where already $N=2$ basis functions approximately capture the characteristic shape of the reference $\mu_\Lambda(h)$ and additional improvements are obtained for $N=3$.

Importantly, even with just a single basis function meaningful approximations of the actual dynamics can be achieved for an adequate choice of the basis.
This becomes conspicuous by comparing the mean peak distances $\mu_\Lambda(h)$ in the case $N=1$ for a $\cosh$-derived basis and the piecewise constant basis in Fig.~\ref{fig:compare_basis_flat} (light green and blue lines).
With only one basis function, the naive averaging cannot be used to approximate the system's dynamics, whereas for the hyperbolic cosine basis a single basis function is sufficient to capture the geometry-dependent characteristics of the system.
This improvement is achieved by leveraging the \textit{a priori} knowledge about the system's sensitivity to the bulk geometry and respecting this sensitivity in the projection step.

\begin{figure}
    \centering
    \includegraphics[width=\columnwidth]{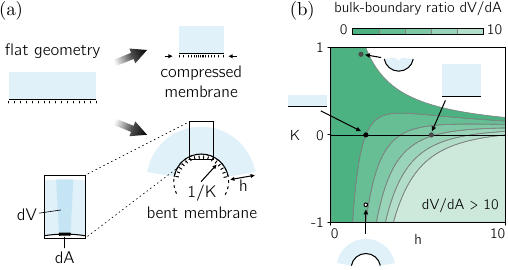}
    \caption{(a)~Schematic visualization of two distinct types of deformation: top right, deformation by compression of the domain along the tangential direction of the membrane, indicated by smaller distance between black stripes. Bottom right, deformation by bending the membrane with curvature $K$ while keeping the metric on the membrane constant, indicated by equidistant black stripes. Inset, definition of the bulk--boundary ratio $\mathrm dV/\mathrm dA$.
    (b)~Bulk--boundary ratio for varying bulk heights and membrane curvatures. For studying the projection method in deforming geometries, we consider a geometry that oscillates between flat (black dot) and outward bent (open dot). Additional characteristic geometries are indicated by gray dots. For large bulk heights and strong inward curvature, the bulk--boundary ratio is ill-defined due to self-intersection of the bulk domain (white area). 
    }
    \label{fig:bulk_boundary_ratio}
\end{figure}
\subsection{Curved and dynamic geometry}\label{sec:applications_deforming}
So far, we discussed the accuracy of the projection ansatz for a flat and static geometry.
How does this method perform in a dynamically deforming geometry?
In general, two types of deformations can be distinguished: deformations that lead to a (local) in-plane compression or stretching of the boundary domain while keeping the shape of the entire domain unchanged, and deformations that keep the metric of the membrane invariant but cause out-of-plane shape changes [Fig.~\ref{fig:bulk_boundary_ratio}a].
In the former case, the membrane is directly affected by the deformation, making it difficult to disentangle the effects of the bulk deformation from the membrane deformation.
Instead, we here focus on the latter case where the reaction--diffusion dynamics on the membrane are not directly affected by the geometry deformation but only indirectly via the coupling to the dynamic bulk.
This allows to compare how well a certain basis choice can capture bulk deformations.

To study such deforming geometries, we apply the projection method presented in Section~\ref{sec:method:deforming} to the toy model in Eq.~\eqref{eq:III:toy_model} that shows height-dependent interrupted coarsening.
For this, we choose a geometry parametrization $\vec r(s,z; t)$ that leaves the metric on the boundary invariant (i.e., $\left| \partial_s \vec r(s,0; t) \right| = 1$) but has a spatio-temporally varying membrane curvature $K(s,t)$ [Fig.~\ref{fig:bulk_boundary_ratio}b, Appendix~\ref{app:geometry}].
Due to this dynamic curvature, the local ratio of bulk volume to membrane area and thereby the magnitude of the bulk field $c(\vec r, t)$ now also vary in space and time~\cite{Thalmeier.etal2016}.
Note that in the static and flat case discussed above this ratio was only affected by the bulk height $h$, since in this case the curvature is $K = 0$ at all times [Fig.~\ref{fig:bulk_boundary_ratio}b].

\begin{figure}
    \centering
    \includegraphics[width=\columnwidth]{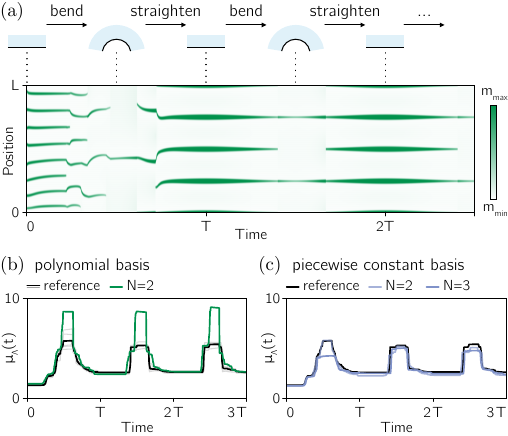}
    \caption{(a)~Sample kymograph of the membrane concentration $m(s,t)$ for the toy model system in Eq.~\eqref{eq:III:toy_model} in a periodically deforming domain with membrane curvature $K(s,t) = K_\text{max}/ 2 \cdot (1 {-} \cos(2 \pi\, t/T ))$ as indicated by the sketches above the kymograph. The length of the membrane in all deformation states is $L=10$, the bulk height is $h=1$. At strongly bent states, the bulkd--boundary ratio is large and the ensuing low concentration of the bulk field $c$ leads to fewer peaks compared to the flat state.
    (b)~Mean peak distance $\mu_\Lambda(t)$ for the reference system (black) and for a projection using a polynomial basis $v_k=\{1, (h{-}z)^2\}$ (green/dark gray), averaged over five samples with different initial conditions each with system size $L=50$. For the reference system, results from individual samples are indicated in gray. The deformation is identical to the one shown in panel (a).
    (c)~Mean peak distance $\mu_\Lambda(t)$ for the reference system (black) and for piecewise constant basis with $N=2$ (light blue/light gray) and $N=3$ (dark blue/gray). For $N=1$, no pattern forms for this bulk height (not shown).
    }
    \label{fig:bending_kymograph}
\end{figure}

Similar to the static case, where the bulk height affects the mean peak distance $\mu_\Lambda(h)$, the interrupted coarsening in a dynamic geometry depends on the bulk--boundary ratio $\mathrm d V/\mathrm dA$ and thus on the membrane curvature $K$ as shown in Fig.~\ref{fig:bulk_boundary_ratio}b.
Modulating the membrane curvature over time thus provides a straightforward method for probing the ability of the projection method to capture effects of the dynamic geometry by comparing the mean peak distance $\mu_\Lambda\bigl(K(t)\bigr)$.
For this, we vary the curvature periodically and homogeneously,
\begin{equation}
    K(s,t) = \frac{K_\text{max}}{2} \bigl(1 - \cos(2 \pi\, t/T ) \bigr)
\end{equation}
with periodicity $T=10^6$ and maximum membrane curvature $K_\text{max} = -1$.
Figure~\ref{fig:bending_kymograph}a shows a typical kymograph of the membrane concentration $m(s,t)$ during multiple deformation cycles.
When the deformation is strongest, the bulk--boundary ratio is maximal (equivalent to an increased bulk height $h$, Fig.~\ref{fig:compare_basis_flat}) and the mean distance between peaks is large.
As the geometry returns to the flat shape, the pattern adapts by spawning additional peaks and thereby decreasing the mean peak distance $\mu_\Lambda$.

In Fig.~\ref{fig:bending_kymograph}b,c we compare the periodic coarsening and spawning of peaks for various basis choices, using a bulk height $h=1$ and maximum curvature $K_\text{max}=-1$.
Similar to the flat case, the naive projection using a single basis function $v_k(z) = \{1\}$ fails to capture the height-dependent dynamics and predicts no patterns at any deformation state at this bulk height (not shown).
To reproduce the reference dynamics, at least two basis functions are required.
Note that a hyperbolic cosine basis yields non-algebraic contributions for the projected PDEs in deforming geometries and is therefore not suitable for numerical implementation.
Instead, we achieve qualitatively matching results by expanding the hyperbolic cosine to lowest order, in this case using a polynomial basis with $v_k = \{1, (h{-}z)^2\}$ [Fig.~\ref{fig:bending_kymograph}b].
For sufficiently small deformations, the polynomial basis captures the reference results with high accuracy, but considerable deviations are observed at strong deformations.
As an alternative, a piecewise constant basis with two or more basis functions may be used [Fig.~\ref{fig:bending_kymograph}c].
Interestingly, the piecwise constant basis captures the pattern statistics more robustly across the entire deformation cycle than the polynomial basis even for only $N=2$ basis functions.
The reason for this lies in the polynomial basis being derived from the steady state concentration profile in a \emph{flat} geometry.
For strong deformations, this steady state profile may change its shape significantly so that the polynomial basis does not capture the relevant bulk gradients anymore.

\section{Discussion and Conclusion}
In conclusion, we have presented a method for using \textit{a priori} knowledge about the physical characteristics of a bulk--boundary reaction--diffusion system to reduce the complexity of the underlying model and thereby simplify the numerical implementation of the system.
Specifically, we proposed a systematic approach to project the field dynamics in a volume onto the surface while respecting the (possibly deforming) geometry of the volume and accounting for spatial gradients normal to the surface.
We showed that this method yields more accurate approximations of the actual dynamics than a naive projection based on averaging the field dynamics over the volume.
At the same time, the projection approach effectively lowers the dimension of the problem by one and thus reduces the computational cost of numerically solving the reaction--diffusion dynamics considerably (approximately 5-fold compared to direct FEM implementations).
As a proof of concept, we applied our method to a generic model showing interrupted coarsening, where the statistical properties of the coarsening dynamics are sensitive to the geometry of the volume.
We evaluated the accuracy of multiple projection methods in both static and deforming geometries and found that for linear bulk reactions a set of two basis functions already leads to good approximations of the actual dynamics for all bulk heights.
Using three or more basis functions provides minor improvements that in general do not outweigh the increase in computation time due to the additional fields.

We emphasize that the projection of the bulk dynamics onto the membrane as presented in this article is only an approximation of the actual dynamics.
In contrast to finite element methods, where accuracy can often be increased simply by refining the mesh~\cite{Brenner.Scott2008}, the accuracy of the projection strongly depends on the basis choice.
Leveraging the full potential of the projection method therefore requires to make use of \textit{a priori} knowledge about the physical properties of the system at hand, such as the system's steady state profiles~\cite{Frey.Brauns2022}.
A physics-informed projection can, by virtue of the dimensionality reduction, simplify the analytical and numerical assessment of pattern-forming system.
However, we stress that no significant benefits compared to FEM implementations are to be expected for uninformed or arbitrary basis choices.

In our derivations of the projection method in deforming geometries we relied on the geometry to be parametrizable and to have a constant height (as measured from the membrane).
As a natural extension to our results, it would be interesting to release these constraints.
In particular, the projection method could be coupled to phase--field implementations of reaction--diffusion dynamics on deforming surfaces, which have been extensively studied in the past~\cite{Ayton.etal2005, Poulsen.Voorhees2018, Valizadeh.Rabczuk2019}.
For this, a key challenge will be to implement a spatio-temporally varying support of the basis functions to account for variable volume height.
Other extensions to our approach could include hydrodynamic coupling of the field dynamics to the deforming geometry~\cite{Mietke.etal2018}, which is an important aspect of many intracellular reaction--diffusion systems~\cite{Goehring.etal2011, Bois.etal2011, Nishikawa.etal2017, Klughammer.etal2018, Illukkumbura.etal2020}, or basis functions that depend on time or are non-local in time, as recently done for a phase-separating system with bulk--boundary coupling~\cite{Caballero.etal2023}.

Beyond the modeling of reaction--diffusion dynamics in parametrized deforming geometries, we expect our results to have valuable applications for systems that reciprocally couple the geometry deformation to the field dynamics, specifically for curvature-generating protein systems~\cite{Alimohamadi.Rangamani2018, Tozzi.etal2019}.
Although the projection method only yields an approximation of the true dynamics, we believe that our approach can have significant advantages for large-scale sampling of the model parameter space in simulations.

\section{Acknowledgements}
We thank Tobias Roth and Henrik Weyer for stimulating discussions.
T.B.~acknowledges support by the Joachim Herz Foundation.
This work was funded by the European Union (ERC, CellGeom, project number 101097810) and the Chan-Zuckerberg Initiative (CZI).

\appendix
\section{General derivation of projection methods}\label{app:derivation}
In this section, we provide a step-by-step derivation of the projection method for arbitrarily deforming geometries.
For additional information on finite element methods, which are conceptually similar to the projection method introduced here, we refer to pertinent textbooks~\cite{Zienkiewicz.etal2005, Brenner.Scott2008}.

We start from a reaction--diffusion equation for a single field in a deforming parametrizable geometry (specified by $\vec r(\vec s,z,t)$, c.f. Fig.~\ref{fig:geometry}, with the bulk extending in $z$-direction) with Robin boundary conditions at the surface $\mathcal S$ (specified by $\vec r_\mathcal{S}(\vec s,t) = \vec r(\vec s,0, t)$) and no-flux boundaries elsewhere:
\begin{align}
    \frac{1}{\sqrt{g}} \partial_t \bigl(\sqrt{g} \, c\bigr) &= D_c \, \Delta_\text{LB} c + f(c) \, , \label{eq:app:pde} \\
    -D_c \left. \vec{\hat n}(\vec s, t) \cdot \vec \nabla c \right|_\mathcal{S} &= f_\mathcal{S}(\left. c \right|_\mathcal{S}; \vec s,t) \,,  \\
    \text{with} \qquad \Delta_\text{LB} c &= \frac{1}{\sqrt{g}} \partial_i \bigl( \sqrt{g} g^{ij} \partial_j c \bigr) \, .
\end{align}
The functions $f$ and $f_\mathcal{S}$ denote the bulk reactions and the bulkd--boundary coupling, respectively, and are both assumed to be polynomial in $c$.
We constrain ourselves to systems with constant bulk height $h$ measured in the normal direction at each point of the membrane.
Consequently, the $z$-direction is chosen to be normal to the membrane.
This allows to write the metric tensor in a block-diagonal form,
\begin{equation}
    g_{ij} = \begin{pmatrix}
        g_{\mathcal S, i'j'} & 0 \\
        0 & 1
    \end{pmatrix} \, , \label{app:eq:metric}
\end{equation}
where the indices $i,j$ run over all coordinates $\{\vec s, z\}$ and the indices $i',j'$ run only over the surface coordinates $\{\vec s\}$ so that $g_{\mathcal S, i'j'}$ is the metric along the coordinates $\vec s$ of the reactive surface $\mathcal{S}$.
In this parametrization, the metric determinant $\det (g_{ij})$ is identical to the determinant of the sub-metric, $g\equiv\det (g_{\mathcal S, i'j'})$.

Analogous to finite element methods~\cite{Brenner.Scott2008}, we start by transforming the PDE for $c(\vec s, z, t)$ to its weak form by introducing a test function $v(z)$ and an inner product $\langle \cdot, \cdot \rangle^{ij}$ on the dynamic geometry along the $z$-direction anchored at each point $\vec s$ on the reactive surface:
\begin{align}
    \langle c(\vec s, z, t) , v(z) \rangle^{ij} &= \int_0^z\!\!\!\mathrm d z \; \sqrt{g}\, g^{ij} \, c(\vec s, z, t) \cdot v(z) \, , \\
    \langle c(\vec s, z, t) , v(z) \rangle &= \int_0^z\!\!\!\mathrm d z \; \sqrt{g}\, c(\vec s, z, t) \cdot v(z) \, ,
\end{align}
where we make use of the special form of the metric~\eqref{app:eq:metric} to single out the $z,z$ component of the inner product $\langle \cdot, \cdot \rangle^{zz} = \langle \cdot, \cdot \rangle$ for easier readability.
With this, the individual terms of the partial differential equation~\eqref{eq:app:pde} in weak form can be expressed as
\begin{widetext}
\begin{align}
     \left\langle \frac{1}{\sqrt{g}} \partial_t \bigl(\sqrt{g} \, c\bigr),v \right\rangle &=  \int_0^z\!\!\!\mathrm d z \; \partial_t \bigl(\sqrt{g} \, c\bigr) \cdot v = \left\langle\partial_t c, v\right\rangle + \left\langle  \frac{1}{\sqrt{g}} c \cdot \partial_t \sqrt{g}, v\right\rangle \, ,\\
     \left\langle D_c \, \Delta_\text{LB} c ,v \right \rangle &= D_c \int_0^z\!\!\!\mathrm d z \; \left[ \partial_{i'} \left( \sqrt{g} \, g^{i'j'} \, \partial_{j'} c\right) \cdot v + \partial_z \left( \sqrt{g} \, \partial_{z} c\right) \cdot v \right] \nonumber \\ 
     &= D_c \, \partial_{i'} \left\langle \partial_{j'} c , v \right\rangle^{i'j'} + D_c  \left[ \sqrt{g} \,  \partial_z c \cdot v \right]_0^h - D_c  \langle \partial_z c, \partial_z v \rangle \, .
\end{align}
\end{widetext}
Note that the index $i'$ runs over all coordinates excluding the $z$-coordinate, so that the derivative $\partial_{i'}$ and the integral over $z$ can be interchanged.
In the next step, we apply a Galerkin projection~\cite{Brenner.Scott2008}: instead of solving the equations for arbitrary fields $c(\vec s, z, t) \in L^2(\mathcal B)$, we only aim to find solutions for a subspace $V \subset L^2(\mathcal S) \times H^1([0,h])$.
We choose this subspace such that all functions $u(\vec s, z, t) \in V$ can be expressed as
\begin{equation}
    u(\vec s, z, t) = u_k(\vec s, t) \, v_k(z) \, , \label{eq:app:separation_ansatz}
\end{equation}
where $v_k(z)$ is a set of $N$ basis functions in the Sobelov space $H^1([0,h])$ with $k \in \{1,\ldots,N{-}1\}$, and we use Einstein summation over double indices.
To highlight the difference between indices denoting spatial coordinates ($i,j$) and indices denoting elements of the basis functions ($k,l,m$), we use co-/contravariant notation for the spatial coordinates but consistently lower indices for the basis functions.
For all elements of the subspace $V$, the weak form of the differential equations can be rewritten further via this separation ansatz~\eqref{eq:app:separation_ansatz}:
\begin{align}
     \left\langle \frac{1}{\sqrt{g}} \partial_t \bigl(\sqrt{g} \, c\bigr),v \right\rangle &= (\partial_t  u_k)  \langle v_k, v \rangle + u_k \, \partial_t \langle v_k, v \rangle \, , \label{eq:app:weak_time} \\
     \left\langle D_c \, \Delta_\text{LB} c ,v \right\rangle &= D_c \, \partial_{i'} \left[ \bigl(\partial_{j'} u_k \bigr) \, \langle v_k, v \rangle^{i'j'} \right] + \nonumber \\
     & \qquad + D_c  \left[ \sqrt{g} \,  \partial_z c \cdot v \right]_0^h  + \nonumber \\
     & \qquad - D_c \, u_k \langle \partial_z v_k, \partial_z v \rangle \, . 
\end{align}
For the boundary term $[\sqrt{g} \, \partial_z c \cdot v]_0^h$, one may at this point insert the boundary conditions $f_\mathcal{S}$ (for $h=0$) and the no-flux boundary condition (for $h=z$).
In the Galerkin projection, the polynomial bulk reaction term $f(c)$ takes a special form.
For the first and second order, for example, the corresponding expressions are
\begin{align}
    \langle c , v \rangle &= u_k \, \langle v_k, v \rangle \, , \\
    \langle c^2, v \rangle &= u_k \, u_l \int_0^z\!\!\!\mathrm d z \; \sqrt{g} \, v_k \, v_l \, v \, ,  \label{eq:app:weak_nonlinear_reaction}
\end{align}
and similarly for higher orders.
Since the weak forms Eqs.~\eqref{eq:app:weak_time}-\eqref{eq:app:weak_nonlinear_reaction} hold for all test functions $v \in V$, one may conveniently choose the basis functions $v_k$ to obtain PDEs for the coefficients $u_k(\vec s, t)$.
Introducing a generalized Gram tensor $G_{kl}^{ij}$, a multilinear variant of the Gram tensor $G^{ij}_{k_1\cdots k_n}$, and a generalized stiffness tensor $A_{kl}^{ij}$ as
\begin{align}
    G_{kl}^{ij} &= \int_0^z\!\!\!\mathrm d z \; \sqrt{g} \, g^{ij} \, v_k \, v_l  = \langle v_k , v_l \rangle^{ij} \, , \\
    G_{k_1\cdots k_n}^{ij} &= \int_0^z\!\!\!\mathrm d z \; \sqrt{g} \, g^{ij} \, v_{k_1} \cdots v_{k_n} \, , \\
    A_{kl}^{ij} &= \int_0^z\!\!\!\mathrm d z \; \sqrt{g} \, g^{ij} \, (\partial_z v_k) (\partial_z v_l) \nonumber \\
    &= \langle \partial_z v_k , \partial_z v_l \rangle^{ij} \, , \label{eq:app:stiffness_matrix}
\end{align}
and abbreviating the entry $G^{zz}_{kl} = \langle v_k, v_l \rangle \equiv G_{kl}$ (similarly for $A_{kl}$ and $G_{k_1\cdots k_n}$) yields a concise formulation for the dynamics of $u_k(\vec s, t)$:
\begin{widetext}
\begin{equation}
    (\partial_t u_k) \, G_{kl} +\underbrace{u_k \, \partial_t G_{kl}\vphantom{\sum_n}}_{g_\text{Geom}} = \underbrace{D_c \, \partial_{i'} \left[ (\partial_{j'} u_k) G^{i'j'}_{kl} \right]\vphantom{\sum_n}}_{g_{\mathcal S\text{-Diff}}} -\underbrace{\left. \sqrt{g} \, f_\mathcal{S}(u_k \, v_k) \cdot v_l \right|_0\vphantom{\sum_n}}_{g_\text{BC}} - \underbrace{D_c \, u_k \, A_{kl}\vphantom{\sum_n}}_{g_{z\text{-Diff}}} + \underbrace{\sum_n \tfrac{1}{n!} \tfrac{\partial^n \! f}{\partial c^n} u_k \, u_{l_2} \cdots u_{l_n} \, G_{kl_2\cdots l_n}}_{g_\text{React}} \, . \label{app:eq:full_projection}
\end{equation}
\end{widetext}
In the case of purely linear bulk reactions, the reaction term reduces to $g_\text{React} = \lambda \, u_k \, G_{kl}$ with linear reaction rate $\lambda$, as applied in the toy model in Section~\ref{sec:applications}.

Importantly, for a parametrizable geometry, the generalized Gram tensor and stiffness tensor can be calculated independently of the actual reaction--diffusion dynamics.
In particular, they can be calculated prior to solving the partial differential equations for the coefficients $u_k$.
The task of solving the bulk dynamics is therefore largely shifted to calculating the Gram and stiffness tensors, which greatly reduces the number of degrees of freedom required to numerically solving the PDEs.
It should be emphasized that this is the primary distinguishing aspect compared to finite element methods: in FEM implementations, the Galerkin projection is performed for each mesh point in the bulk and on the boundary~\cite{Brenner.Scott2008}.
For a mesh size $a$ and system width $L$ and height $h$ in $d$ dimensions, this results in a matrix of size $\mathcal{O}\bigl((L/a)^{d-1} \, (h/a) \bigr)$ that needs to be diagonalized in each step.
In contrast, by projecting the bulk dynamics onto the membrane using $N$ basis functions the computational complexity is reduced to diagonalizing a matrix of size $\mathcal{O}\bigl(N \, (L/a)^{d{-}1} \bigr)$, where FEM (or other standard methods) can be used to solve the dynamics on the membrane.
Specifically, the projection method changes the size of the relevant matrices by a factor $N \cdot a/h$, so that the computational benefit is largest for low mesh sizes $a$ and a small number of basis functions $N$.

This improvement is possible by explicitly allowing highly nonlinear basis functions in the projection approach and by adapting the basis functions choice to the physical problem that is to be solved.
For FEM, in contrast, it is often convenient to choose basis functions $v_k$ with small support centered around the mesh points, as this renders the Gram matrix $G$ as well as other important matrices sparse and thereby speeds up numerical applications.
The projection method abolishes this need for basis functions with small support in the bulk, since the Gram and stiffness matrices for the bulk dynamics only need to be calculated once before solving the system dynamics (however, Gram and stiffness matrix for the boundary dynamics still need to diagonalized at each time step).
When choosing the basis for the projection method, one is therefore not constrained by the small support of the bulk basis functions, and instead the basis can be tuned to the physical properties of the system.
A good choice of basis functions is therefore crucial for leveraging the full potential of this dimensionality reduction approach.

\section{Basis of orthogonal step functions}\label{app:step}
In this section, we discuss how the stiffness matrix $A_{kl}$ needs to be adjusted when choosing a basis consisting of step functions.
For this, consider a one-dimensional bulk $[0,h]$ subdivided into $N$ segments at $\{0, z_1, z_2, \ldots, z_{N-1}, h \}$ so that each segment has size $h_k = z_{k{+}1}-z_k$.
The corresponding basis functions are 
\begin{equation}
    v_k = \begin{cases}
        1 & z_k \leq z \leq z_{k+1} \,, \\
        0 & \text{else}.
    \end{cases}
\end{equation}
This basis is orthogonal by design and thus the Gram matrix takes the form $G = \text{diag}(\{h_0, \ldots, h_{N{-}1} \})$.
However, to construct the stiffness matrix $A_{kl}$ it is necessary to find derivatives of the basis functions or at least derivatives in the weak sense~\cite{Brenner.Scott2008}, i.e., functions $w_k(z)$ for which
\begin{equation}
    \int_0^h \!\! \mathrm d z \; v_k(z) \cdot (\partial_z \phi(z)) = -  \int_0^h \!\! \mathrm d z \; w_k(z) \cdot \phi(z) 
\end{equation}
for all differentiable test functions $\phi(z)$.
Since step functions are not weakly differentiable~\cite{Adams.Fournier2003}, the stiffness matrix $A_{kl}$ as defined in Eq.~\eqref{eq:app:stiffness_matrix} is not applicable for a step function basis.

At first glance, one may try to circumvent this issue by approximating the step functions using smooth analogues, for example a logistic sigmoid $\sigma_w(z) = (1+e^{-z/w})^{-1}$, where the parameter $w$ quantifies the width of the step.
The basis elements may then be written as $v_k=\sigma_w(z-z_k) \, \sigma_w(z_{k{+}1} -z)$.
This basis is not orthogonal anymore, however the Gram basis remains to lowest order $G = \text{diag}(\{h_0, \ldots, h_{N{-}1} \}) + \mathcal{O}(w)$.
The stiffness matrix, on the other hand, takes a tridiagonal form to lowest order,
\begin{equation}
    A = \frac{1}{6w} \begin{pmatrix}
        -2 & 1 & 0 &  \\
        1 & -2 & 1 & \cdots \\
        0 & 1 & -2 &  \\
         & \vdots &  & \ddots 
    \end{pmatrix} + \mathcal{O}(w) \, .
\end{equation}
As a result, the diffusion term $g_\text{Diff}$ with such a stiffness matrix enters the differential equation for the coefficients (e.g., Eq.~\eqref{eq:II.9}) with a prefactor ${\sim}w^{-1}$, leading to diverging terms in the limit $w\to 0$.
This is because the true gradients $\partial_z c(z)$ are poorly approximated by the step functions, where gradients are $\partial_z u \approx (u_{k{+}1} - u_k)/w$ in the $w$-neighborhood of the nodes $z_k$ and zero elsewhere.
To be able to use step functions (or their smooth analogues) as a basis, the stiffness matrix needs to be reconstructed artificially.

For this, it is necessary to define approximations of the gradient terms $\partial_z \tilde v_k$ as stand-ins for the weak derivatives of the actual basis $v_k$.
As the simplest approach, one may emulate standard FEM strategies and use linear interpolation between the separate layers [Fig.~XX].
This yields piecewise constant functions for the approximate gradients:
\begin{equation}
     \partial_z \tilde v_k = \begin{cases}
        \frac{2}{h_{k{-}1} + h_k} & z_k - \frac{h_{k{-}1}}{2} \leq z \leq z_k + \frac{h_k}{2} \\
        \frac{2}{h_{k} + h_{k{+}1}} & z_{k{+}1} - \frac{h_{k}}{2} \leq z \leq z_{k{+}1} - \frac{h_{k+1}}{2} \\
        0 & \text{else} \, .
        \end{cases}
\end{equation}%
The resulting reconstructed stiffness matrix $\bar A_{kl}$ then evaluates to the tridiagonal form
\begin{align}
    \bar A_{kk} &= \frac{2}{h_k + h_{k{+}1}}+ \frac{2}{h_k + h_{k{-}1}} \,, \nonumber \\
    \bar A_{k,k{\pm}1} &= -\frac{2}{h_k + h_{k{\pm}1}} \,, \label{eq:app:reconstructed_stiffness_matrix}
\end{align}
and zero for all other entries.
\textit{A priori} knowledge about the system, e.g., about the steady state distribution, can be used to obtain better estimates for the gradients.
For all results presented in this paper we use the FEM-inspired stiffness matrix as defined in Eq.~\eqref{eq:app:reconstructed_stiffness_matrix} when working with the step function basis.

\section{Numerical implementation}%
\begin{table}[!t]
\renewcommand{\arraystretch}{1.3}%
    \centering
\begin{tabularx}{\columnwidth}{|l|r|X|}
    \hline
    \textbf{Parameter} & \textbf{Value} & \textbf{Description}\\\hline
    $D_{c}$ & $1$ & Bulk diffusion \\ \hline
    $D_{m}$ & $10^{-4}$ & Membrane diffusion \\ \hline
    $\lambda$ & $1$ & Bulk degradation rate \\ \hline 
    $\epsilon$ & $10^{-2}$ & Membrane degradation rate \\ \hline
    $n_0$ & $1$ & Average initial mass on membrane \\ \hline
    $L$ & $100$ & System width \\ \hline
    $\ell = \sqrt{D_c/\lambda}$ & $1$ & Characteristic length scale \\ \hline 
    $p$ & $1.1$ & Effective membrane activation rate \\ \hline
\end{tabularx}
    \caption{List of parameters used in simulations.}
    \label{tab:params}
\end{table}%
To test the projection method with the toy model as presented in Section~\ref{sec:applications}, we performed numerical simulations using COMSOL Multiphysics~v6.0 for the different projection approaches (basis choices) and, for reference, for a non-projected system, with simulation parameters as stated in Table~\ref{tab:params}.
For the reference simulations, the bulk dynamics were solved in a two-dimensional domain coupled to a one-dimensional boundary accounting for the membrane dynamics via reactive boundary conditions.
The boundary domain was meshed regularly with a mesh size of $0.05$.
The bulk domain was subsequently meshed using a triangular Delauney tesselation (irregular) combined with local adaptive mesh refinement, where the mesh resolution was reduced at distances far away from the membrane controlled by setting the software's configuration parameter \texttt{Maximum element growth rate} to $1.5$.
Note that the adaptive mesh refinement already assumes shallow gradients far away from the membrane.
For comparison, we also produced solutions for a bulk domain that was meshed in a regular rectangular grid with mesh size $0.05$.
For the projected systems, all dynamics (membrane and projected bulk) were solved on a one-dimensional domain corresponding to the boundary in the reference simulations.
This domain was meshed regularly with a mesh size of $0.05$.

\begin{figure}
    \centering
    \includegraphics[width=\columnwidth]{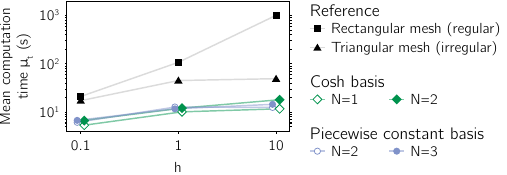}
    \caption{Benchmark of solution times for the toy model in a flat geometry for a system with explicit bulk dynamics (reference, black, with regular rectangular mesh (squares) and local adaptive triangular mesh (triangles)), for a projected system with basis functions derived from the hyperbolic cosine steady state concentration profile (green/dark gray, diamonds), and for a projected system with piecewise constant basis functions (blue/light gray, circles). Horizontal displacement is added to ensure that data points are distinguishable but does not correspond to different bulk heights $h$. Data points represent the average value over five samples each. Relative standard deviations ($\sigma_t/\mu_t$) are $<0.45$ (reference, $h=10$) and $< 0.22$ for all other values. Lines are added to guide the eye.}
    \label{fig:app:benchmark}
\end{figure}
In Fig.~\ref{fig:app:benchmark} we compare the solution times for different simulation methods (reference and projections) for three different values of the system height $h$.
Each data point in Fig.~\ref{fig:app:benchmark} is the mean of five samples obtained from different initial conditions and otherwise identical parameters.
In the reference simulations (both regular mesh and adaptive refined mesh, black), solution times are consistently larger than ten seconds, and increase to approximately one minute (adaptive refined mesh) and ten minutes (regular mesh) at large system heights $h=10$ where the number of mesh points becomes large.
In contrast, for simulations using bulk projection with up to $N=3$ basis functions the solution time remained below 20~seconds, and reached approximately five seconds for low system heights ($h=0.1$, where coarsening is interrupted earlier and a steady state is reached faster, c.f.~Fig.~\ref{fig:kymograph}).
Importantly, the number of mesh points in the projected system is independent of the bulk height in contrast to the reference system, and therefore the computational speedup is most prominent for large bulk height and a regular mesh.
Comparing the reference simulations with adaptive refined mesh to the simulations using $N=1$ basis functions derived from a hyperbolic cosine, a speedup factor of ${>}3$ is achieved for $h =0.1$, and a factor ${>}4$ for $h = 1$ and $h = 10$.
For benchmarking the solution times, the simulations were performed on a 64~core AMD~Ryzen Threadripper processor at 2.70~GHz.

\section{Projection of mass-conserving systems}
In reaction--diffusion systems, the diffusion part always conserves the total protein mass $n = \int \! \mathrm d V \, c(\vec x, t)$, i.e., for a system with no-flux boundaries at all boundaries and without any reaction terms defined by
\begin{equation}
    \partial_t c(\vec x, t) = D_c \, \nabla^2 c(\vec x, t)
\end{equation}
the change in total mass is~\cite{Burkart.etal2022}
\begin{equation}
    \partial_t n = \int_\mathcal{B} \!\! \mathrm d V \; \partial_t c(\vec x, t) = \left.  \nabla c(\vec x, t) \right|_{\partial \mathcal B} = 0 \, .
\end{equation}
This property of mass-conserving diffusion should be preserved under the projection method, in particular for systems where all reactions are also mass-conserving.
Under what circumstances does the projection method respect mass conservation in the diffusion?

For this, it is sufficient to consider diffusion in a one-dimensional bulk (i.e., no diffusion parallel to the membrane) with no-flux boundary conditions at both boundaries.
For such a system, the governing equations for the fields $u_k(t)$ derived in Eq.~\eqref{app:eq:full_projection} reduce to
\begin{equation}
    G_{kl} \, (\partial_t u_k) = -D_c \, u_k \, A_{kl} \, .
\end{equation}
Without loss of generality, assume furthermore that the basis $\{v_k\}$ is orthonormal so that the Gram matrix is a unit matrix $G_{kl} = \delta_{kl}$.
Since each component $k$ contributes to the total concentration by $u_k \, v_k(z)$, the change in the total mass is then
\begin{equation}
    \partial_t n = \partial_t u_k \, \int \! \mathrm d z \; v_k(z) = - D_c  \left[\int \! \mathrm d z \; v_k(z)\right] \cdot A_{kl} \, u_l \, .
\end{equation}
The total mass is thus conserved if and only if
\begin{equation}
    A_{kl} \, \int \! \mathrm d z \; v_k(z) = 0 \quad \forall \,  l \, . \label{app:eq:mass_conservation}
\end{equation}
This poses an additional constraint on the basis choice, complementing the requirements to approximate the true distribution $c(z, t)$ and its gradient $\partial_z c(z,t)$ well.
We now show that this requirement is always fulfilled if one of the basis functions is constant in $z$-direction.
For this, assume that $v_0(z) = 1/\sqrt{h}$ in a bulk domain of height $h$.
This allows to rewrite the integral over a single basis function in terms of the Gram matrix,
\begin{equation}
    \int \! \mathrm d z \; v_k(z) = \sqrt{h} \int \! \mathrm d z \; v_k(z) \, v_0(z) = \sqrt{h} \, G_{k0} \, .
\end{equation}
Inserting this into the requirement for mass conservation in Eq.~\eqref{app:eq:mass_conservation} and using the orthonormality of the basis $G_{kl} = \delta_{kl}$ yields
\begin{equation}
    \sqrt{h} A_{kl} \, G_{k0} = \sqrt{h} A_{0l} = 0 \, , 
\end{equation}
which is always fulfilled for a basis that contains a constant basis function (in this case $v_0$).
This follows from the fact that $A_{kl} = \langle \partial_z v_k , \, \partial_z v_l \rangle$, and by choice of the basis function $\partial_z v_0 = 0$.

Since the orthonormalization step only affects the representation of the system but not the dynamics, the same line of argument holds for a basis where only a linear combination of basis functions is constant, i.e., where 
\begin{equation}
    \alpha_k \, \partial_z v_k(z) = 0
\end{equation}
for real constants $\alpha_k$.
In particular, this includes a set of piecewise constant basis functions as presented in Sec.~\ref{sec:applications}.

\section{Parametrization of evolving geometries}\label{app:geometry}
In this section, we specify the parametrization used to evaluate the projection method in a deforming geometry as discussed in Sec.~\ref{sec:applications_deforming}.
Since we use that the basis functions $v_k(z)$ are identical for all points $\vec r(\vec s, 0)$ on the membrane $\mathcal{S}$, a necessary constraint for the parametrizations is that they have constant height $h$ everywhere.
Note that here the height is defined as the distance from the membrane $\mathcal{S}$ to the opposing no-flux boundary in the direction normal to $\mathcal S$.
This also implies that these two boundaries are always parallel.
In addition, deformations need to avoid intersections of the bulk with itself, since in such cases the metric diverges.
Therefore, the maximum height of the bulk in a deforming geometry is constrained by the maximum (principal) curvature, $h < 1/K_\text{max}$.
Furthermore, we here choose a parametrization for which the metric on the membrane is constant and unity, $\left. g_{ij} \right|_{z=0} = \delta_{ij}$, since this ensures that no stretching or compression of the membrane can affect the coarsening dynamics and all observed effects are due to the bulk dynamics.

 For the examples in Sec.~\ref{sec:applications_deforming}, we examine a geometry consisting of an annular segment with spatially constant but temporally varying curvature $K(t)$.
This geometry is parametrized by
\begin{equation}
    \vec r(s, z, t) = \frac{1}{K(t)} \begin{pmatrix}
        \bigl( 1 + z \, K(t) \bigr) \sin (s \, K(t)) \\
        \bigl( 1 + z \, K(t) \bigr) \cos (s \, K(t)) -1
    \end{pmatrix} \, , \label{eq:app:parametrization_bending}
\end{equation}
where we choose $K(t) = K_\text{max} \cdot \frac{1}{2} \left((1 - \cos (\frac{2 \pi \,t}{T})\right)$ with $K_\text{max} < 0$.
This geometry is flat at $t = n \cdot 2 \pi \, T$ for $n \in \mathbb{N}$ and morphs into an arc with curvature $|K_\text{max}|$ at $t = (2 n+1) \cdot \pi \,  T$ with a periodicity $T$.
This parametrization ensures that the surface is never inward bent and therefore no problematic self-intersections can occur.

\bibliography{bibliography.bib}

\begin{thebibliography}{67}%
\makeatletter
\providecommand \@ifxundefined [1]{%
 \@ifx{#1\undefined}
}%
\providecommand \@ifnum [1]{%
 \ifnum #1\expandafter \@firstoftwo
 \else \expandafter \@secondoftwo
 \fi
}%
\providecommand \@ifx [1]{%
 \ifx #1\expandafter \@firstoftwo
 \else \expandafter \@secondoftwo
 \fi
}%
\providecommand \natexlab [1]{#1}%
\providecommand \enquote  [1]{``#1''}%
\providecommand \bibnamefont  [1]{#1}%
\providecommand \bibfnamefont [1]{#1}%
\providecommand \citenamefont [1]{#1}%
\providecommand \href@noop [0]{\@secondoftwo}%
\providecommand \href [0]{\begingroup \@sanitize@url \@href}%
\providecommand \@href[1]{\@@startlink{#1}\@@href}%
\providecommand \@@href[1]{\endgroup#1\@@endlink}%
\providecommand \@sanitize@url [0]{\catcode `\\12\catcode `\$12\catcode `\&12\catcode `\#12\catcode `\^12\catcode `\_12\catcode `\%12\relax}%
\providecommand \@@startlink[1]{}%
\providecommand \@@endlink[0]{}%
\providecommand \url  [0]{\begingroup\@sanitize@url \@url }%
\providecommand \@url [1]{\endgroup\@href {#1}{\urlprefix }}%
\providecommand \urlprefix  [0]{URL }%
\providecommand \Eprint [0]{\href }%
\providecommand \doibase [0]{https://doi.org/}%
\providecommand \selectlanguage [0]{\@gobble}%
\providecommand \bibinfo  [0]{\@secondoftwo}%
\providecommand \bibfield  [0]{\@secondoftwo}%
\providecommand \translation [1]{[#1]}%
\providecommand \BibitemOpen [0]{}%
\providecommand \bibitemStop [0]{}%
\providecommand \bibitemNoStop [0]{.\EOS\space}%
\providecommand \EOS [0]{\spacefactor3000\relax}%
\providecommand \BibitemShut  [1]{\csname bibitem#1\endcsname}%
\let\auto@bib@innerbib\@empty
\bibitem [{\citenamefont {Wolpert}(1981)}]{Wolpert1981}%
  \BibitemOpen
  \bibfield  {author} {\bibinfo {author} {\bibfnamefont {L.}~\bibnamefont {Wolpert}},\ }\href {https://doi.org/10.1098/rstb.1981.0152} {\bibfield  {journal} {\bibinfo  {journal} {Philosophical Transactions of the Royal Society of London. B, Biological Sciences}\ }\textbf {\bibinfo {volume} {295}},\ \bibinfo {pages} {441} (\bibinfo {year} {1981})}\BibitemShut {NoStop}%
\bibitem [{\citenamefont {Desai}\ and\ \citenamefont {Kapral}(2009)}]{Desai.Kapral2009}%
  \BibitemOpen
  \bibfield  {author} {\bibinfo {author} {\bibfnamefont {R.~C.}\ \bibnamefont {Desai}}\ and\ \bibinfo {author} {\bibfnamefont {R.}~\bibnamefont {Kapral}},\ }\href@noop {} {\emph {\bibinfo {title} {{Dynamics of Self-Organized and Self-Assembled Structures}}}}\ (\bibinfo  {publisher} {Cambridge University Press},\ \bibinfo {year} {2009})\BibitemShut {NoStop}%
\bibitem [{\citenamefont {Burkart}\ \emph {et~al.}(2022)\citenamefont {Burkart}, \citenamefont {Wigbers}, \citenamefont {Würthner},\ and\ \citenamefont {Frey}}]{Burkart.etal2022}%
  \BibitemOpen
  \bibfield  {author} {\bibinfo {author} {\bibfnamefont {T.}~\bibnamefont {Burkart}}, \bibinfo {author} {\bibfnamefont {M.~C.}\ \bibnamefont {Wigbers}}, \bibinfo {author} {\bibfnamefont {L.}~\bibnamefont {Würthner}},\ and\ \bibinfo {author} {\bibfnamefont {E.}~\bibnamefont {Frey}},\ }\href {https://doi.org/10.1038/s42254-022-00461-3} {\bibfield  {journal} {\bibinfo  {journal} {Nature Reviews Physics}\ }\textbf {\bibinfo {volume} {4}},\ \bibinfo {pages} {511} (\bibinfo {year} {2022})}\BibitemShut {NoStop}%
\bibitem [{\citenamefont {Raskin}\ and\ \citenamefont {Boer}(1999)}]{Raskin.Boer1999}%
  \BibitemOpen
  \bibfield  {author} {\bibinfo {author} {\bibfnamefont {D.~M.}\ \bibnamefont {Raskin}}\ and\ \bibinfo {author} {\bibfnamefont {P.~A. J.~d.}\ \bibnamefont {Boer}},\ }\href {https://doi.org/10.1073/pnas.96.9.4971} {\bibfield  {journal} {\bibinfo  {journal} {Proceedings of the National Academy of Sciences}\ }\textbf {\bibinfo {volume} {96}},\ \bibinfo {pages} {4971} (\bibinfo {year} {1999})}\BibitemShut {NoStop}%
\bibitem [{\citenamefont {Ramm}\ \emph {et~al.}(2019)\citenamefont {Ramm}, \citenamefont {Heermann},\ and\ \citenamefont {Schwille}}]{Ramm.etal2019}%
  \BibitemOpen
  \bibfield  {author} {\bibinfo {author} {\bibfnamefont {B.}~\bibnamefont {Ramm}}, \bibinfo {author} {\bibfnamefont {T.}~\bibnamefont {Heermann}},\ and\ \bibinfo {author} {\bibfnamefont {P.}~\bibnamefont {Schwille}},\ }\href {https://doi.org/10.1007/s00018-019-03218-x} {\bibfield  {journal} {\bibinfo  {journal} {Cellular and Molecular Life Sciences}\ }\textbf {\bibinfo {volume} {76}},\ \bibinfo {pages} {4245} (\bibinfo {year} {2019})}\BibitemShut {NoStop}%
\bibitem [{\citenamefont {Chiou}\ \emph {et~al.}(2016)\citenamefont {Chiou}, \citenamefont {Balasubramanian},\ and\ \citenamefont {Lew}}]{Chiou.etal2016}%
  \BibitemOpen
  \bibfield  {author} {\bibinfo {author} {\bibfnamefont {J.-G.}\ \bibnamefont {Chiou}}, \bibinfo {author} {\bibfnamefont {M.~K.}\ \bibnamefont {Balasubramanian}},\ and\ \bibinfo {author} {\bibfnamefont {D.~J.}\ \bibnamefont {Lew}},\ }\href {https://doi.org/10.1146/annurev-cellbio-100616-060856} {\bibfield  {journal} {\bibinfo  {journal} {Annual Review of Cell and Developmental Biology}\ }\textbf {\bibinfo {volume} {33}},\ \bibinfo {pages} {1} (\bibinfo {year} {2016})}\BibitemShut {NoStop}%
\bibitem [{\citenamefont {Motegi}\ and\ \citenamefont {Seydoux}(2013)}]{Motegi.Seydoux2013}%
  \BibitemOpen
  \bibfield  {author} {\bibinfo {author} {\bibfnamefont {F.}~\bibnamefont {Motegi}}\ and\ \bibinfo {author} {\bibfnamefont {G.}~\bibnamefont {Seydoux}},\ }\href {https://doi.org/10.1098/rstb.2013.0010} {\bibfield  {journal} {\bibinfo  {journal} {Philosophical Transactions of the Royal Society B: Biological Sciences}\ }\textbf {\bibinfo {volume} {368}},\ \bibinfo {pages} {20130010} (\bibinfo {year} {2013})}\BibitemShut {NoStop}%
\bibitem [{\citenamefont {Walani}\ \emph {et~al.}(2014)\citenamefont {Walani}, \citenamefont {Torres},\ and\ \citenamefont {Agrawal}}]{Walani.etal2014}%
  \BibitemOpen
  \bibfield  {author} {\bibinfo {author} {\bibfnamefont {N.}~\bibnamefont {Walani}}, \bibinfo {author} {\bibfnamefont {J.}~\bibnamefont {Torres}},\ and\ \bibinfo {author} {\bibfnamefont {A.}~\bibnamefont {Agrawal}},\ }\href {https://doi.org/10.1103/physreve.89.062715} {\bibfield  {journal} {\bibinfo  {journal} {Physical Review E}\ }\textbf {\bibinfo {volume} {89}},\ \bibinfo {pages} {062715} (\bibinfo {year} {2014})}\BibitemShut {NoStop}%
\bibitem [{\citenamefont {Rogez}\ \emph {et~al.}(2019)\citenamefont {Rogez}, \citenamefont {Würthner}, \citenamefont {Petrova}, \citenamefont {Zierhut}, \citenamefont {Saczko-Brack}, \citenamefont {Huergo}, \citenamefont {Batters}, \citenamefont {Frey},\ and\ \citenamefont {Veigel}}]{Rogez.etal2019}%
  \BibitemOpen
  \bibfield  {author} {\bibinfo {author} {\bibfnamefont {B.}~\bibnamefont {Rogez}}, \bibinfo {author} {\bibfnamefont {L.}~\bibnamefont {Würthner}}, \bibinfo {author} {\bibfnamefont {A.~B.}\ \bibnamefont {Petrova}}, \bibinfo {author} {\bibfnamefont {F.~B.}\ \bibnamefont {Zierhut}}, \bibinfo {author} {\bibfnamefont {D.}~\bibnamefont {Saczko-Brack}}, \bibinfo {author} {\bibfnamefont {M.-A.}\ \bibnamefont {Huergo}}, \bibinfo {author} {\bibfnamefont {C.}~\bibnamefont {Batters}}, \bibinfo {author} {\bibfnamefont {E.}~\bibnamefont {Frey}},\ and\ \bibinfo {author} {\bibfnamefont {C.}~\bibnamefont {Veigel}},\ }\href {https://doi.org/10.1038/s41467-019-11268-9} {\bibfield  {journal} {\bibinfo  {journal} {Nature Communications}\ }\textbf {\bibinfo {volume} {10}},\ \bibinfo {pages} {3305} (\bibinfo {year} {2019})}\BibitemShut {NoStop}%
\bibitem [{\citenamefont {Mahapatra}\ \emph {et~al.}(2021)\citenamefont {Mahapatra}, \citenamefont {Saintillan},\ and\ \citenamefont {Rangamani}}]{Mahapatra.etal2021}%
  \BibitemOpen
  \bibfield  {author} {\bibinfo {author} {\bibfnamefont {A.}~\bibnamefont {Mahapatra}}, \bibinfo {author} {\bibfnamefont {D.}~\bibnamefont {Saintillan}},\ and\ \bibinfo {author} {\bibfnamefont {P.}~\bibnamefont {Rangamani}},\ }\href {https://doi.org/10.1039/d1sm00502b} {\bibfield  {journal} {\bibinfo  {journal} {Soft Matter}\ }\textbf {\bibinfo {volume} {17}},\ \bibinfo {pages} {8373} (\bibinfo {year} {2021})}\BibitemShut {NoStop}%
\bibitem [{\citenamefont {Roux}\ \emph {et~al.}(2021)\citenamefont {Roux}, \citenamefont {Tozzi}, \citenamefont {Walani}, \citenamefont {Quiroga}, \citenamefont {Zalvidea}, \citenamefont {Trepat}, \citenamefont {Staykova}, \citenamefont {Arroyo},\ and\ \citenamefont {Roca-Cusachs}}]{Roux.etal2021}%
  \BibitemOpen
  \bibfield  {author} {\bibinfo {author} {\bibfnamefont {A.-L.~L.}\ \bibnamefont {Roux}}, \bibinfo {author} {\bibfnamefont {C.}~\bibnamefont {Tozzi}}, \bibinfo {author} {\bibfnamefont {N.}~\bibnamefont {Walani}}, \bibinfo {author} {\bibfnamefont {X.}~\bibnamefont {Quiroga}}, \bibinfo {author} {\bibfnamefont {D.}~\bibnamefont {Zalvidea}}, \bibinfo {author} {\bibfnamefont {X.}~\bibnamefont {Trepat}}, \bibinfo {author} {\bibfnamefont {M.}~\bibnamefont {Staykova}}, \bibinfo {author} {\bibfnamefont {M.}~\bibnamefont {Arroyo}},\ and\ \bibinfo {author} {\bibfnamefont {P.}~\bibnamefont {Roca-Cusachs}},\ }\href {https://doi.org/10.1038/s41467-021-26591-3} {\bibfield  {journal} {\bibinfo  {journal} {Nature Communications}\ }\textbf {\bibinfo {volume} {12}},\ \bibinfo {pages} {6550} (\bibinfo {year} {2021})}\BibitemShut {NoStop}%
\bibitem [{\citenamefont {Würthner}\ \emph {et~al.}(2023)\citenamefont {Würthner}, \citenamefont {Goychuk},\ and\ \citenamefont {Frey}}]{Wuerthner.etal2023}%
  \BibitemOpen
  \bibfield  {author} {\bibinfo {author} {\bibfnamefont {L.}~\bibnamefont {Würthner}}, \bibinfo {author} {\bibfnamefont {A.}~\bibnamefont {Goychuk}},\ and\ \bibinfo {author} {\bibfnamefont {E.}~\bibnamefont {Frey}},\ }\href {https://doi.org/10.1103/physreve.108.014404} {\bibfield  {journal} {\bibinfo  {journal} {Physical Review E}\ }\textbf {\bibinfo {volume} {108}},\ \bibinfo {pages} {014404} (\bibinfo {year} {2023})},\ \Eprint {https://arxiv.org/abs/2205.02820} {2205.02820} \BibitemShut {NoStop}%
\bibitem [{\citenamefont {Hubatsch}\ \emph {et~al.}(2019)\citenamefont {Hubatsch}, \citenamefont {Peglion}, \citenamefont {Reich}, \citenamefont {Rodrigues}, \citenamefont {Hirani}, \citenamefont {Illukkumbura},\ and\ \citenamefont {Goehring}}]{Hubatsch.etal2019}%
  \BibitemOpen
  \bibfield  {author} {\bibinfo {author} {\bibfnamefont {L.}~\bibnamefont {Hubatsch}}, \bibinfo {author} {\bibfnamefont {F.}~\bibnamefont {Peglion}}, \bibinfo {author} {\bibfnamefont {J.~D.}\ \bibnamefont {Reich}}, \bibinfo {author} {\bibfnamefont {N.~T.~L.}\ \bibnamefont {Rodrigues}}, \bibinfo {author} {\bibfnamefont {N.}~\bibnamefont {Hirani}}, \bibinfo {author} {\bibfnamefont {R.}~\bibnamefont {Illukkumbura}},\ and\ \bibinfo {author} {\bibfnamefont {N.~W.}\ \bibnamefont {Goehring}},\ }\href {https://doi.org/10.1038/s41567-019-0601-x} {\bibfield  {journal} {\bibinfo  {journal} {Nature Physics}\ }\textbf {\bibinfo {volume} {15}},\ \bibinfo {pages} {1078} (\bibinfo {year} {2019})}\BibitemShut {NoStop}%
\bibitem [{\citenamefont {Marée}\ \emph {et~al.}(2012)\citenamefont {Marée}, \citenamefont {Grieneisen},\ and\ \citenamefont {Edelstein-Keshet}}]{Maree.etal2012}%
  \BibitemOpen
  \bibfield  {author} {\bibinfo {author} {\bibfnamefont {A.~F.~M.}\ \bibnamefont {Marée}}, \bibinfo {author} {\bibfnamefont {V.~A.}\ \bibnamefont {Grieneisen}},\ and\ \bibinfo {author} {\bibfnamefont {L.}~\bibnamefont {Edelstein-Keshet}},\ }\href {https://doi.org/10.1371/journal.pcbi.1002402} {\bibfield  {journal} {\bibinfo  {journal} {PLoS Computational Biology}\ }\textbf {\bibinfo {volume} {8}},\ \bibinfo {pages} {e1002402} (\bibinfo {year} {2012})}\BibitemShut {NoStop}%
\bibitem [{\citenamefont {Camley}\ \emph {et~al.}(2017)\citenamefont {Camley}, \citenamefont {Zhao}, \citenamefont {Li}, \citenamefont {Levine},\ and\ \citenamefont {Rappel}}]{Camley.etal2017}%
  \BibitemOpen
  \bibfield  {author} {\bibinfo {author} {\bibfnamefont {B.~A.}\ \bibnamefont {Camley}}, \bibinfo {author} {\bibfnamefont {Y.}~\bibnamefont {Zhao}}, \bibinfo {author} {\bibfnamefont {B.}~\bibnamefont {Li}}, \bibinfo {author} {\bibfnamefont {H.}~\bibnamefont {Levine}},\ and\ \bibinfo {author} {\bibfnamefont {W.-J.}\ \bibnamefont {Rappel}},\ }\href {https://doi.org/10.1103/physreve.95.012401} {\bibfield  {journal} {\bibinfo  {journal} {Physical Review E}\ }\textbf {\bibinfo {volume} {95}},\ \bibinfo {pages} {012401} (\bibinfo {year} {2017})},\ \Eprint {https://arxiv.org/abs/1609.01764} {1609.01764} \BibitemShut {NoStop}%
\bibitem [{\citenamefont {Fu}\ \emph {et~al.}(2023)\citenamefont {Fu}, \citenamefont {Burkart}, \citenamefont {Maryshev}, \citenamefont {Franquelim}, \citenamefont {Merino-Salomón}, \citenamefont {Reverte-López}, \citenamefont {Frey},\ and\ \citenamefont {Schwille}}]{Fu.etal2023}%
  \BibitemOpen
  \bibfield  {author} {\bibinfo {author} {\bibfnamefont {M.}~\bibnamefont {Fu}}, \bibinfo {author} {\bibfnamefont {T.}~\bibnamefont {Burkart}}, \bibinfo {author} {\bibfnamefont {I.}~\bibnamefont {Maryshev}}, \bibinfo {author} {\bibfnamefont {H.~G.}\ \bibnamefont {Franquelim}}, \bibinfo {author} {\bibfnamefont {A.}~\bibnamefont {Merino-Salomón}}, \bibinfo {author} {\bibfnamefont {M.}~\bibnamefont {Reverte-López}}, \bibinfo {author} {\bibfnamefont {E.}~\bibnamefont {Frey}},\ and\ \bibinfo {author} {\bibfnamefont {P.}~\bibnamefont {Schwille}},\ }\href {https://doi.org/10.1038/s41567-023-02058-8} {\bibfield  {journal} {\bibinfo  {journal} {Nature Physics}\ }\textbf {\bibinfo {volume} {19}},\ \bibinfo {pages} {1211} (\bibinfo {year} {2023})}\BibitemShut {NoStop}%
\bibitem [{\citenamefont {Miller}\ \emph {et~al.}(2018)\citenamefont {Miller}, \citenamefont {Stoop},\ and\ \citenamefont {Dunkel}}]{Miller.etal2018}%
  \BibitemOpen
  \bibfield  {author} {\bibinfo {author} {\bibfnamefont {P.~W.}\ \bibnamefont {Miller}}, \bibinfo {author} {\bibfnamefont {N.}~\bibnamefont {Stoop}},\ and\ \bibinfo {author} {\bibfnamefont {J.}~\bibnamefont {Dunkel}},\ }\href {https://doi.org/10.1103/physrevlett.120.268001} {\bibfield  {journal} {\bibinfo  {journal} {Physical Review Letters}\ }\textbf {\bibinfo {volume} {120}},\ \bibinfo {pages} {268001} (\bibinfo {year} {2018})},\ \Eprint {https://arxiv.org/abs/1710.02247} {1710.02247} \BibitemShut {NoStop}%
\bibitem [{\citenamefont {Litschel}\ \emph {et~al.}(2018)\citenamefont {Litschel}, \citenamefont {Ramm}, \citenamefont {Maas}, \citenamefont {Heymann},\ and\ \citenamefont {Schwille}}]{Litschel.etal2018}%
  \BibitemOpen
  \bibfield  {author} {\bibinfo {author} {\bibfnamefont {T.}~\bibnamefont {Litschel}}, \bibinfo {author} {\bibfnamefont {B.}~\bibnamefont {Ramm}}, \bibinfo {author} {\bibfnamefont {R.}~\bibnamefont {Maas}}, \bibinfo {author} {\bibfnamefont {M.}~\bibnamefont {Heymann}},\ and\ \bibinfo {author} {\bibfnamefont {P.}~\bibnamefont {Schwille}},\ }\href {https://doi.org/10.1002/anie.201808750} {\bibfield  {journal} {\bibinfo  {journal} {Angewandte Chemie International Edition}\ }\textbf {\bibinfo {volume} {57}},\ \bibinfo {pages} {16286} (\bibinfo {year} {2018})}\BibitemShut {NoStop}%
\bibitem [{\citenamefont {Tamemoto}\ and\ \citenamefont {Noguchi}(2020)}]{Tamemoto.Noguchi2020}%
  \BibitemOpen
  \bibfield  {author} {\bibinfo {author} {\bibfnamefont {N.}~\bibnamefont {Tamemoto}}\ and\ \bibinfo {author} {\bibfnamefont {H.}~\bibnamefont {Noguchi}},\ }\href {https://doi.org/10.1038/s41598-020-76695-x} {\bibfield  {journal} {\bibinfo  {journal} {Scientific Reports}\ }\textbf {\bibinfo {volume} {10}},\ \bibinfo {pages} {19582} (\bibinfo {year} {2020})},\ \Eprint {https://arxiv.org/abs/2007.10606} {2007.10606} \BibitemShut {NoStop}%
\bibitem [{\citenamefont {Hörning}\ and\ \citenamefont {Shibata}(2019)}]{Hoerning.Shibata2019}%
  \BibitemOpen
  \bibfield  {author} {\bibinfo {author} {\bibfnamefont {M.}~\bibnamefont {Hörning}}\ and\ \bibinfo {author} {\bibfnamefont {T.}~\bibnamefont {Shibata}},\ }\href {https://doi.org/10.1016/j.bpj.2018.12.012} {\bibfield  {journal} {\bibinfo  {journal} {Biophysical Journal}\ }\textbf {\bibinfo {volume} {116}},\ \bibinfo {pages} {372} (\bibinfo {year} {2019})}\BibitemShut {NoStop}%
\bibitem [{\citenamefont {Brinkmann}\ \emph {et~al.}(2018)\citenamefont {Brinkmann}, \citenamefont {Mercker}, \citenamefont {Richter},\ and\ \citenamefont {Marciniak-Czochra}}]{Brinkmann.etal2018}%
  \BibitemOpen
  \bibfield  {author} {\bibinfo {author} {\bibfnamefont {F.}~\bibnamefont {Brinkmann}}, \bibinfo {author} {\bibfnamefont {M.}~\bibnamefont {Mercker}}, \bibinfo {author} {\bibfnamefont {T.}~\bibnamefont {Richter}},\ and\ \bibinfo {author} {\bibfnamefont {A.}~\bibnamefont {Marciniak-Czochra}},\ }\href {https://doi.org/10.1371/journal.pcbi.1006259} {\bibfield  {journal} {\bibinfo  {journal} {PLOS Computational Biology}\ }\textbf {\bibinfo {volume} {14}},\ \bibinfo {pages} {e1006259} (\bibinfo {year} {2018})}\BibitemShut {NoStop}%
\bibitem [{\citenamefont {Mietke}\ \emph {et~al.}(2018)\citenamefont {Mietke}, \citenamefont {Jülicher},\ and\ \citenamefont {Sbalzarini}}]{Mietke.etal2018}%
  \BibitemOpen
  \bibfield  {author} {\bibinfo {author} {\bibfnamefont {A.}~\bibnamefont {Mietke}}, \bibinfo {author} {\bibfnamefont {F.}~\bibnamefont {Jülicher}},\ and\ \bibinfo {author} {\bibfnamefont {I.~F.}\ \bibnamefont {Sbalzarini}},\ }\href {https://doi.org/10.1073/pnas.1810896115} {\bibfield  {journal} {\bibinfo  {journal} {Proceedings of the National Academy of Sciences}\ }\textbf {\bibinfo {volume} {116}},\ \bibinfo {pages} {29} (\bibinfo {year} {2018})}\BibitemShut {NoStop}%
\bibitem [{\citenamefont {Meacci}\ \emph {et~al.}(2006)\citenamefont {Meacci}, \citenamefont {Ries}, \citenamefont {Fischer-Friedrich}, \citenamefont {Kahya}, \citenamefont {Schwille},\ and\ \citenamefont {Kruse}}]{Meacci.etal2006}%
  \BibitemOpen
  \bibfield  {author} {\bibinfo {author} {\bibfnamefont {G.}~\bibnamefont {Meacci}}, \bibinfo {author} {\bibfnamefont {J.}~\bibnamefont {Ries}}, \bibinfo {author} {\bibfnamefont {E.}~\bibnamefont {Fischer-Friedrich}}, \bibinfo {author} {\bibfnamefont {N.}~\bibnamefont {Kahya}}, \bibinfo {author} {\bibfnamefont {P.}~\bibnamefont {Schwille}},\ and\ \bibinfo {author} {\bibfnamefont {K.}~\bibnamefont {Kruse}},\ }\href {https://doi.org/10.1088/1478-3975/3/4/003} {\bibfield  {journal} {\bibinfo  {journal} {Physical Biology}\ }\textbf {\bibinfo {volume} {3}},\ \bibinfo {pages} {255} (\bibinfo {year} {2006})},\ \Eprint {https://arxiv.org/abs/q-bio/0701046} {q-bio/0701046} \BibitemShut {NoStop}%
\bibitem [{\citenamefont {Dziuk}\ and\ \citenamefont {Elliott}(2007)}]{Dziuk.Elliott2007}%
  \BibitemOpen
  \bibfield  {author} {\bibinfo {author} {\bibfnamefont {G.}~\bibnamefont {Dziuk}}\ and\ \bibinfo {author} {\bibfnamefont {C.~M.}\ \bibnamefont {Elliott}},\ }\href {https://doi.org/10.1093/imanum/drl023} {\bibfield  {journal} {\bibinfo  {journal} {IMA Journal of Numerical Analysis}\ }\textbf {\bibinfo {volume} {27}},\ \bibinfo {pages} {262} (\bibinfo {year} {2007})}\BibitemShut {NoStop}%
\bibitem [{\citenamefont {Madzvamuse}\ \emph {et~al.}(2015)\citenamefont {Madzvamuse}, \citenamefont {Chung},\ and\ \citenamefont {Venkataraman}}]{Madzvamuse.etal2015}%
  \BibitemOpen
  \bibfield  {author} {\bibinfo {author} {\bibfnamefont {A.}~\bibnamefont {Madzvamuse}}, \bibinfo {author} {\bibfnamefont {A.~H.~W.}\ \bibnamefont {Chung}},\ and\ \bibinfo {author} {\bibfnamefont {C.}~\bibnamefont {Venkataraman}},\ }\href {https://doi.org/10.1098/rspa.2014.0546} {\bibfield  {journal} {\bibinfo  {journal} {Proceedings of the Royal Society A: Mathematical, Physical and Engineering Sciences}\ }\textbf {\bibinfo {volume} {471}},\ \bibinfo {pages} {20140546} (\bibinfo {year} {2015})},\ \Eprint {https://arxiv.org/abs/1501.05769} {1501.05769} \BibitemShut {NoStop}%
\bibitem [{\citenamefont {Chen}\ and\ \citenamefont {Ling}(2020)}]{Chen.Ling2020}%
  \BibitemOpen
  \bibfield  {author} {\bibinfo {author} {\bibfnamefont {M.}~\bibnamefont {Chen}}\ and\ \bibinfo {author} {\bibfnamefont {L.}~\bibnamefont {Ling}},\ }\href {https://doi.org/10.1016/j.jcp.2019.109166} {\bibfield  {journal} {\bibinfo  {journal} {Journal of Computational Physics}\ }\textbf {\bibinfo {volume} {405}},\ \bibinfo {pages} {109166} (\bibinfo {year} {2020})},\ \Eprint {https://arxiv.org/abs/1906.01370} {1906.01370} \BibitemShut {NoStop}%
\bibitem [{\citenamefont {Bergdorf}\ \emph {et~al.}(2010)\citenamefont {Bergdorf}, \citenamefont {Sbalzarini},\ and\ \citenamefont {Koumoutsakos}}]{Bergdorf.etal2010}%
  \BibitemOpen
  \bibfield  {author} {\bibinfo {author} {\bibfnamefont {M.}~\bibnamefont {Bergdorf}}, \bibinfo {author} {\bibfnamefont {I.~F.}\ \bibnamefont {Sbalzarini}},\ and\ \bibinfo {author} {\bibfnamefont {P.}~\bibnamefont {Koumoutsakos}},\ }\href {https://doi.org/10.1007/s00285-009-0315-2} {\bibfield  {journal} {\bibinfo  {journal} {Journal of Mathematical Biology}\ }\textbf {\bibinfo {volume} {61}},\ \bibinfo {pages} {649} (\bibinfo {year} {2010})}\BibitemShut {NoStop}%
\bibitem [{\citenamefont {Hansbo}\ \emph {et~al.}(2016)\citenamefont {Hansbo}, \citenamefont {Larson},\ and\ \citenamefont {Zahedi}}]{Hansbo.etal2016}%
  \BibitemOpen
  \bibfield  {author} {\bibinfo {author} {\bibfnamefont {P.}~\bibnamefont {Hansbo}}, \bibinfo {author} {\bibfnamefont {M.~G.}\ \bibnamefont {Larson}},\ and\ \bibinfo {author} {\bibfnamefont {S.}~\bibnamefont {Zahedi}},\ }\href {https://doi.org/10.1016/j.cma.2016.04.012} {\bibfield  {journal} {\bibinfo  {journal} {Computer Methods in Applied Mechanics and Engineering}\ }\textbf {\bibinfo {volume} {307}},\ \bibinfo {pages} {96} (\bibinfo {year} {2016})},\ \Eprint {https://arxiv.org/abs/1502.07142} {1502.07142} \BibitemShut {NoStop}%
\bibitem [{\citenamefont {Suchde}\ and\ \citenamefont {Kuhnert}(2019)}]{Suchde.Kuhnert2019}%
  \BibitemOpen
  \bibfield  {author} {\bibinfo {author} {\bibfnamefont {P.}~\bibnamefont {Suchde}}\ and\ \bibinfo {author} {\bibfnamefont {J.}~\bibnamefont {Kuhnert}},\ }\href {https://doi.org/10.1016/j.jcp.2019.06.031} {\bibfield  {journal} {\bibinfo  {journal} {Journal of Computational Physics}\ }\textbf {\bibinfo {volume} {395}},\ \bibinfo {pages} {38} (\bibinfo {year} {2019})},\ \Eprint {https://arxiv.org/abs/1902.08107} {1902.08107} \BibitemShut {NoStop}%
\bibitem [{\citenamefont {Wendland}\ and\ \citenamefont {Künemund}(2020)}]{Wendland.Kuenemund2020}%
  \BibitemOpen
  \bibfield  {author} {\bibinfo {author} {\bibfnamefont {H.}~\bibnamefont {Wendland}}\ and\ \bibinfo {author} {\bibfnamefont {J.}~\bibnamefont {Künemund}},\ }\href {https://doi.org/10.1007/s10444-020-09803-0} {\bibfield  {journal} {\bibinfo  {journal} {Advances in Computational Mathematics}\ }\textbf {\bibinfo {volume} {46}},\ \bibinfo {pages} {64} (\bibinfo {year} {2020})}\BibitemShut {NoStop}%
\bibitem [{\citenamefont {Lervåg}\ and\ \citenamefont {Lowengrub}(2015)}]{Lervag.Lowengrub2015}%
  \BibitemOpen
  \bibfield  {author} {\bibinfo {author} {\bibfnamefont {K.~Y.}\ \bibnamefont {Lervåg}}\ and\ \bibinfo {author} {\bibfnamefont {J.}~\bibnamefont {Lowengrub}},\ }\href {https://doi.org/10.4310/cms.2015.v13.n6.a6} {\bibfield  {journal} {\bibinfo  {journal} {Communications in Mathematical Sciences}\ }\textbf {\bibinfo {volume} {13}},\ \bibinfo {pages} {1473} (\bibinfo {year} {2015})},\ \Eprint {https://arxiv.org/abs/1407.7480} {1407.7480} \BibitemShut {NoStop}%
\bibitem [{\citenamefont {Loose}\ \emph {et~al.}(2008)\citenamefont {Loose}, \citenamefont {Fischer-Friedrich}, \citenamefont {Ries}, \citenamefont {Kruse},\ and\ \citenamefont {Schwille}}]{Loose.etal2008}%
  \BibitemOpen
  \bibfield  {author} {\bibinfo {author} {\bibfnamefont {M.}~\bibnamefont {Loose}}, \bibinfo {author} {\bibfnamefont {E.}~\bibnamefont {Fischer-Friedrich}}, \bibinfo {author} {\bibfnamefont {J.}~\bibnamefont {Ries}}, \bibinfo {author} {\bibfnamefont {K.}~\bibnamefont {Kruse}},\ and\ \bibinfo {author} {\bibfnamefont {P.}~\bibnamefont {Schwille}},\ }\href {https://doi.org/10.1126/science.1154413} {\bibfield  {journal} {\bibinfo  {journal} {Science}\ }\textbf {\bibinfo {volume} {320}},\ \bibinfo {pages} {789} (\bibinfo {year} {2008})}\BibitemShut {NoStop}%
\bibitem [{\citenamefont {Bement}\ \emph {et~al.}(2015)\citenamefont {Bement}, \citenamefont {Leda}, \citenamefont {Moe}, \citenamefont {Kita}, \citenamefont {Larson}, \citenamefont {Golding}, \citenamefont {Pfeuti}, \citenamefont {Su}, \citenamefont {Miller}, \citenamefont {Goryachev},\ and\ \citenamefont {Dassow}}]{Bement.etal2015}%
  \BibitemOpen
  \bibfield  {author} {\bibinfo {author} {\bibfnamefont {W.~M.}\ \bibnamefont {Bement}}, \bibinfo {author} {\bibfnamefont {M.}~\bibnamefont {Leda}}, \bibinfo {author} {\bibfnamefont {A.~M.}\ \bibnamefont {Moe}}, \bibinfo {author} {\bibfnamefont {A.~M.}\ \bibnamefont {Kita}}, \bibinfo {author} {\bibfnamefont {M.~E.}\ \bibnamefont {Larson}}, \bibinfo {author} {\bibfnamefont {A.~E.}\ \bibnamefont {Golding}}, \bibinfo {author} {\bibfnamefont {C.}~\bibnamefont {Pfeuti}}, \bibinfo {author} {\bibfnamefont {K.-C.}\ \bibnamefont {Su}}, \bibinfo {author} {\bibfnamefont {A.~L.}\ \bibnamefont {Miller}}, \bibinfo {author} {\bibfnamefont {A.~B.}\ \bibnamefont {Goryachev}},\ and\ \bibinfo {author} {\bibfnamefont {G.~v.}\ \bibnamefont {Dassow}},\ }\href {https://doi.org/10.1038/ncb3251} {\bibfield  {journal} {\bibinfo  {journal} {Nature Cell Biology}\ }\textbf {\bibinfo {volume} {17}},\ \bibinfo {pages} {1471} (\bibinfo {year} {2015})}\BibitemShut {NoStop}%
\bibitem [{\citenamefont {Paquin-Lefebvre}\ \emph {et~al.}(2020)\citenamefont {Paquin-Lefebvre}, \citenamefont {Xu}, \citenamefont {DiPietro}, \citenamefont {Lindsay},\ and\ \citenamefont {Jilkine}}]{Paquin-Lefebvre.etal2020}%
  \BibitemOpen
  \bibfield  {author} {\bibinfo {author} {\bibfnamefont {F.}~\bibnamefont {Paquin-Lefebvre}}, \bibinfo {author} {\bibfnamefont {B.}~\bibnamefont {Xu}}, \bibinfo {author} {\bibfnamefont {K.~L.}\ \bibnamefont {DiPietro}}, \bibinfo {author} {\bibfnamefont {A.~E.}\ \bibnamefont {Lindsay}},\ and\ \bibinfo {author} {\bibfnamefont {A.}~\bibnamefont {Jilkine}},\ }\href {https://doi.org/10.1016/j.jtbi.2020.110242} {\bibfield  {journal} {\bibinfo  {journal} {Journal of Theoretical Biology}\ }\textbf {\bibinfo {volume} {497}},\ \bibinfo {pages} {110242} (\bibinfo {year} {2020})},\ \Eprint {https://arxiv.org/abs/1910.10203} {1910.10203} \BibitemShut {NoStop}%
\bibitem [{\citenamefont {Halatek}\ \emph {et~al.}(2018)\citenamefont {Halatek}, \citenamefont {Brauns},\ and\ \citenamefont {Frey}}]{Halatek.etal2018}%
  \BibitemOpen
  \bibfield  {author} {\bibinfo {author} {\bibfnamefont {J.}~\bibnamefont {Halatek}}, \bibinfo {author} {\bibfnamefont {F.}~\bibnamefont {Brauns}},\ and\ \bibinfo {author} {\bibfnamefont {E.}~\bibnamefont {Frey}},\ }\href {https://doi.org/10.1098/rstb.2017.0107} {\bibfield  {journal} {\bibinfo  {journal} {Philosophical Transactions of the Royal Society B: Biological Sciences}\ }\textbf {\bibinfo {volume} {373}},\ \bibinfo {pages} {20170107} (\bibinfo {year} {2018})},\ \Eprint {https://arxiv.org/abs/1802.07169} {1802.07169} \BibitemShut {NoStop}%
\bibitem [{\citenamefont {Brauns}\ \emph {et~al.}(2021{\natexlab{a}})\citenamefont {Brauns}, \citenamefont {Pawlik}, \citenamefont {Halatek}, \citenamefont {Kerssemakers}, \citenamefont {Frey},\ and\ \citenamefont {Dekker}}]{Brauns.etal2021}%
  \BibitemOpen
  \bibfield  {author} {\bibinfo {author} {\bibfnamefont {F.}~\bibnamefont {Brauns}}, \bibinfo {author} {\bibfnamefont {G.}~\bibnamefont {Pawlik}}, \bibinfo {author} {\bibfnamefont {J.}~\bibnamefont {Halatek}}, \bibinfo {author} {\bibfnamefont {J.}~\bibnamefont {Kerssemakers}}, \bibinfo {author} {\bibfnamefont {E.}~\bibnamefont {Frey}},\ and\ \bibinfo {author} {\bibfnamefont {C.}~\bibnamefont {Dekker}},\ }\href {https://doi.org/10.1038/s41467-021-23412-5} {\bibfield  {journal} {\bibinfo  {journal} {Nature Communications}\ }\textbf {\bibinfo {volume} {12}},\ \bibinfo {pages} {3312} (\bibinfo {year} {2021}{\natexlab{a}})}\BibitemShut {NoStop}%
\bibitem [{\citenamefont {Würthner}\ \emph {et~al.}(2022)\citenamefont {Würthner}, \citenamefont {Brauns}, \citenamefont {Pawlik}, \citenamefont {Halatek}, \citenamefont {Kerssemakers}, \citenamefont {Dekker},\ and\ \citenamefont {Frey}}]{Wuerthner.etal2022}%
  \BibitemOpen
  \bibfield  {author} {\bibinfo {author} {\bibfnamefont {L.}~\bibnamefont {Würthner}}, \bibinfo {author} {\bibfnamefont {F.}~\bibnamefont {Brauns}}, \bibinfo {author} {\bibfnamefont {G.}~\bibnamefont {Pawlik}}, \bibinfo {author} {\bibfnamefont {J.}~\bibnamefont {Halatek}}, \bibinfo {author} {\bibfnamefont {J.}~\bibnamefont {Kerssemakers}}, \bibinfo {author} {\bibfnamefont {C.}~\bibnamefont {Dekker}},\ and\ \bibinfo {author} {\bibfnamefont {E.}~\bibnamefont {Frey}},\ }\href {https://doi.org/10.1073/pnas.2206888119} {\bibfield  {journal} {\bibinfo  {journal} {Proceedings of the National Academy of Sciences}\ }\textbf {\bibinfo {volume} {119}},\ \bibinfo {pages} {e2206888119} (\bibinfo {year} {2022})},\ \Eprint {https://arxiv.org/abs/2111.12043} {2111.12043} \BibitemShut {NoStop}%
\bibitem [{\citenamefont {Levine}\ and\ \citenamefont {Rappel}(2005)}]{Levine.Rappel2005}%
  \BibitemOpen
  \bibfield  {author} {\bibinfo {author} {\bibfnamefont {H.}~\bibnamefont {Levine}}\ and\ \bibinfo {author} {\bibfnamefont {W.-J.}\ \bibnamefont {Rappel}},\ }\href {https://doi.org/10.1103/physreve.72.061912} {\bibfield  {journal} {\bibinfo  {journal} {Physical Review E}\ }\textbf {\bibinfo {volume} {72}},\ \bibinfo {pages} {061912} (\bibinfo {year} {2005})}\BibitemShut {NoStop}%
\bibitem [{\citenamefont {Li}\ \emph {et~al.}(2009)\citenamefont {Li}, \citenamefont {Lowengrub}, \citenamefont {Rätz},\ and\ \citenamefont {Voigt}}]{Li.etal2009}%
  \BibitemOpen
  \bibfield  {author} {\bibinfo {author} {\bibfnamefont {X.}~\bibnamefont {Li}}, \bibinfo {author} {\bibfnamefont {J.}~\bibnamefont {Lowengrub}}, \bibinfo {author} {\bibfnamefont {A.}~\bibnamefont {Rätz}},\ and\ \bibinfo {author} {\bibfnamefont {A.}~\bibnamefont {Voigt}},\ }\href {https://doi.org/10.4310/cms.2009.v7.n1.a4} {\bibfield  {journal} {\bibinfo  {journal} {Communications in Mathematical Sciences}\ }\textbf {\bibinfo {volume} {7}},\ \bibinfo {pages} {81} (\bibinfo {year} {2009})}\BibitemShut {NoStop}%
\bibitem [{\citenamefont {Teigen}\ \emph {et~al.}(2009)\citenamefont {Teigen}, \citenamefont {Li}, \citenamefont {Lowengrub}, \citenamefont {Wang},\ and\ \citenamefont {Voigt}}]{Teigen.etal2009}%
  \BibitemOpen
  \bibfield  {author} {\bibinfo {author} {\bibfnamefont {K.~E.}\ \bibnamefont {Teigen}}, \bibinfo {author} {\bibfnamefont {X.}~\bibnamefont {Li}}, \bibinfo {author} {\bibfnamefont {J.}~\bibnamefont {Lowengrub}}, \bibinfo {author} {\bibfnamefont {F.}~\bibnamefont {Wang}},\ and\ \bibinfo {author} {\bibfnamefont {A.}~\bibnamefont {Voigt}},\ }\href {https://doi.org/10.4310/cms.2009.v7.n4.a10} {\bibfield  {journal} {\bibinfo  {journal} {Communications in Mathematical Sciences}\ }\textbf {\bibinfo {volume} {7}},\ \bibinfo {pages} {1009} (\bibinfo {year} {2009})}\BibitemShut {NoStop}%
\bibitem [{\citenamefont {Yu}\ \emph {et~al.}(2012)\citenamefont {Yu}, \citenamefont {Chen},\ and\ \citenamefont {Thornton}}]{Yu.etal2012}%
  \BibitemOpen
  \bibfield  {author} {\bibinfo {author} {\bibfnamefont {H.-C.}\ \bibnamefont {Yu}}, \bibinfo {author} {\bibfnamefont {H.-Y.}\ \bibnamefont {Chen}},\ and\ \bibinfo {author} {\bibfnamefont {K.}~\bibnamefont {Thornton}},\ }\href {https://doi.org/10.1088/0965-0393/20/7/075008} {\bibfield  {journal} {\bibinfo  {journal} {Modelling and Simulation in Materials Science and Engineering}\ }\textbf {\bibinfo {volume} {20}},\ \bibinfo {pages} {075008} (\bibinfo {year} {2012})},\ \Eprint {https://arxiv.org/abs/1107.5341} {1107.5341} \BibitemShut {NoStop}%
\bibitem [{\citenamefont {Marth}\ and\ \citenamefont {Voigt}(2014)}]{Marth.Voigt2014}%
  \BibitemOpen
  \bibfield  {author} {\bibinfo {author} {\bibfnamefont {W.}~\bibnamefont {Marth}}\ and\ \bibinfo {author} {\bibfnamefont {A.}~\bibnamefont {Voigt}},\ }\href {https://doi.org/10.1007/s00285-013-0704-4} {\bibfield  {journal} {\bibinfo  {journal} {Journal of Mathematical Biology}\ }\textbf {\bibinfo {volume} {69}},\ \bibinfo {pages} {91} (\bibinfo {year} {2014})}\BibitemShut {NoStop}%
\bibitem [{\citenamefont {Rätz}\ and\ \citenamefont {Röger}(2014)}]{Raetz.Roeger2014}%
  \BibitemOpen
  \bibfield  {author} {\bibinfo {author} {\bibfnamefont {A.}~\bibnamefont {Rätz}}\ and\ \bibinfo {author} {\bibfnamefont {M.}~\bibnamefont {Röger}},\ }\href {https://doi.org/10.1088/0951-7715/27/8/1805} {\bibfield  {journal} {\bibinfo  {journal} {Nonlinearity}\ }\textbf {\bibinfo {volume} {27}},\ \bibinfo {pages} {1805} (\bibinfo {year} {2014})},\ \Eprint {https://arxiv.org/abs/1305.6172} {1305.6172} \BibitemShut {NoStop}%
\bibitem [{\citenamefont {Abels}\ \emph {et~al.}(2015)\citenamefont {Abels}, \citenamefont {Lam},\ and\ \citenamefont {Stinner}}]{Abels.etal2015}%
  \BibitemOpen
  \bibfield  {author} {\bibinfo {author} {\bibfnamefont {H.}~\bibnamefont {Abels}}, \bibinfo {author} {\bibfnamefont {K.~F.}\ \bibnamefont {Lam}},\ and\ \bibinfo {author} {\bibfnamefont {B.}~\bibnamefont {Stinner}},\ }\href {https://doi.org/10.1137/15m1009093} {\bibfield  {journal} {\bibinfo  {journal} {SIAM Journal on Mathematical Analysis}\ }\textbf {\bibinfo {volume} {47}},\ \bibinfo {pages} {3687} (\bibinfo {year} {2015})}\BibitemShut {NoStop}%
\bibitem [{\citenamefont {Moure}\ and\ \citenamefont {Gomez}(2020)}]{Moure.Gomez2020}%
  \BibitemOpen
  \bibfield  {author} {\bibinfo {author} {\bibfnamefont {A.}~\bibnamefont {Moure}}\ and\ \bibinfo {author} {\bibfnamefont {H.}~\bibnamefont {Gomez}},\ }\href {https://doi.org/10.1039/d0sm00473a} {\bibfield  {journal} {\bibinfo  {journal} {Soft Matter}\ }\textbf {\bibinfo {volume} {16}},\ \bibinfo {pages} {5177} (\bibinfo {year} {2020})}\BibitemShut {NoStop}%
\bibitem [{\citenamefont {Valizadeh}\ and\ \citenamefont {Rabczuk}(2019)}]{Valizadeh.Rabczuk2019}%
  \BibitemOpen
  \bibfield  {author} {\bibinfo {author} {\bibfnamefont {N.}~\bibnamefont {Valizadeh}}\ and\ \bibinfo {author} {\bibfnamefont {T.}~\bibnamefont {Rabczuk}},\ }\href {https://doi.org/10.1016/j.cma.2019.03.043} {\bibfield  {journal} {\bibinfo  {journal} {Computer Methods in Applied Mechanics and Engineering}\ }\textbf {\bibinfo {volume} {351}},\ \bibinfo {pages} {599} (\bibinfo {year} {2019})}\BibitemShut {NoStop}%
\bibitem [{\citenamefont {Biben}\ \emph {et~al.}(2005)\citenamefont {Biben}, \citenamefont {Kassner},\ and\ \citenamefont {Misbah}}]{Biben.etal2005}%
  \BibitemOpen
  \bibfield  {author} {\bibinfo {author} {\bibfnamefont {T.}~\bibnamefont {Biben}}, \bibinfo {author} {\bibfnamefont {K.}~\bibnamefont {Kassner}},\ and\ \bibinfo {author} {\bibfnamefont {C.}~\bibnamefont {Misbah}},\ }\href {https://doi.org/10.1103/physreve.72.041921} {\bibfield  {journal} {\bibinfo  {journal} {Physical Review E}\ }\textbf {\bibinfo {volume} {72}},\ \bibinfo {pages} {041921} (\bibinfo {year} {2005})}\BibitemShut {NoStop}%
\bibitem [{\citenamefont {Rätz}\ and\ \citenamefont {Voigt}(2006)}]{Raetz.Voigt2006}%
  \BibitemOpen
  \bibfield  {author} {\bibinfo {author} {\bibfnamefont {A.}~\bibnamefont {Rätz}}\ and\ \bibinfo {author} {\bibfnamefont {A.}~\bibnamefont {Voigt}},\ }\href {https://doi.org/10.4310/cms.2006.v4.n3.a5} {\bibfield  {journal} {\bibinfo  {journal} {Communications in Mathematical Sciences}\ }\textbf {\bibinfo {volume} {4}},\ \bibinfo {pages} {575} (\bibinfo {year} {2006})}\BibitemShut {NoStop}%
\bibitem [{\citenamefont {Elliott}\ \emph {et~al.}(2011)\citenamefont {Elliott}, \citenamefont {Stinner}, \citenamefont {Styles},\ and\ \citenamefont {Welford}}]{Elliott.etal2011}%
  \BibitemOpen
  \bibfield  {author} {\bibinfo {author} {\bibfnamefont {C.~M.}\ \bibnamefont {Elliott}}, \bibinfo {author} {\bibfnamefont {B.}~\bibnamefont {Stinner}}, \bibinfo {author} {\bibfnamefont {V.}~\bibnamefont {Styles}},\ and\ \bibinfo {author} {\bibfnamefont {R.}~\bibnamefont {Welford}},\ }\href {https://doi.org/10.1093/imanum/drq005} {\bibfield  {journal} {\bibinfo  {journal} {IMA Journal of Numerical Analysis}\ }\textbf {\bibinfo {volume} {31}},\ \bibinfo {pages} {786} (\bibinfo {year} {2011})}\BibitemShut {NoStop}%
\bibitem [{\citenamefont {Poulsen}\ and\ \citenamefont {Voorhees}(2018)}]{Poulsen.Voorhees2018}%
  \BibitemOpen
  \bibfield  {author} {\bibinfo {author} {\bibfnamefont {S.~O.}\ \bibnamefont {Poulsen}}\ and\ \bibinfo {author} {\bibfnamefont {P.~W.}\ \bibnamefont {Voorhees}},\ }\href {https://doi.org/10.1142/s0219876218500147} {\bibfield  {journal} {\bibinfo  {journal} {International Journal of Computational Methods}\ }\textbf {\bibinfo {volume} {15}},\ \bibinfo {pages} {1850014} (\bibinfo {year} {2018})}\BibitemShut {NoStop}%
\bibitem [{\citenamefont {Yu}\ \emph {et~al.}(2020)\citenamefont {Yu}, \citenamefont {Guo},\ and\ \citenamefont {Lowengrub}}]{Yu.etal2020}%
  \BibitemOpen
  \bibfield  {author} {\bibinfo {author} {\bibfnamefont {F.}~\bibnamefont {Yu}}, \bibinfo {author} {\bibfnamefont {Z.}~\bibnamefont {Guo}},\ and\ \bibinfo {author} {\bibfnamefont {J.}~\bibnamefont {Lowengrub}},\ }\href {https://doi.org/10.1016/j.jcp.2019.109174} {\bibfield  {journal} {\bibinfo  {journal} {Journal of Computational Physics}\ }\textbf {\bibinfo {volume} {406}},\ \bibinfo {pages} {109174} (\bibinfo {year} {2020})}\BibitemShut {NoStop}%
\bibitem [{\citenamefont {Zienkiewicz}\ \emph {et~al.}(2005)\citenamefont {Zienkiewicz}, \citenamefont {Taylor},\ and\ \citenamefont {Zhou}}]{Zienkiewicz.etal2005}%
  \BibitemOpen
  \bibfield  {author} {\bibinfo {author} {\bibfnamefont {O.~C.}\ \bibnamefont {Zienkiewicz}}, \bibinfo {author} {\bibfnamefont {R.~L.}\ \bibnamefont {Taylor}},\ and\ \bibinfo {author} {\bibfnamefont {J.~Z.}\ \bibnamefont {Zhou}},\ }\href@noop {} {\emph {\bibinfo {title} {{The finite element method: its basis and fundamentals}}}}\ (\bibinfo  {publisher} {Elsevier},\ \bibinfo {year} {2005})\BibitemShut {NoStop}%
\bibitem [{\citenamefont {Brenner}\ and\ \citenamefont {Scott}(2008)}]{Brenner.Scott2008}%
  \BibitemOpen
  \bibfield  {author} {\bibinfo {author} {\bibfnamefont {S.~C.}\ \bibnamefont {Brenner}}\ and\ \bibinfo {author} {\bibfnamefont {L.~R.}\ \bibnamefont {Scott}},\ }\bibfield  {journal} {\bibinfo  {journal} {Texts in Applied Mathematics}\ }\href {https://doi.org/10.1007/978-0-387-75934-0} {10.1007/978-0-387-75934-0} (\bibinfo {year} {2008})\BibitemShut {NoStop}%
\bibitem [{\citenamefont {Frey}\ and\ \citenamefont {Brauns}(2022)}]{Frey.Brauns2022}%
  \BibitemOpen
  \bibfield  {author} {\bibinfo {author} {\bibfnamefont {E.}~\bibnamefont {Frey}}\ and\ \bibinfo {author} {\bibfnamefont {F.}~\bibnamefont {Brauns}},\ }in\ \href {https://doi.org/10.1093/oso/9780192858313.003.0011} {\emph {\bibinfo {booktitle} {Active Matter and Nonequilibrium Statistical Physics}}},\ \bibinfo {series} {Lecture Notes of the Les Houches Summer School}, Vol.\ \bibinfo {volume} {112},\ \bibinfo {editor} {edited by\ \bibinfo {editor} {\bibfnamefont {J.}~\bibnamefont {Tailleur}}}\ (\bibinfo  {publisher} {Oxford Academic},\ \bibinfo {address} {Oxford},\ \bibinfo {year} {2022})\ pp.\ \bibinfo {pages} {347--445}\BibitemShut {NoStop}%
\bibitem [{\citenamefont {Brauns}\ \emph {et~al.}(2020)\citenamefont {Brauns}, \citenamefont {Halatek},\ and\ \citenamefont {Frey}}]{Brauns.etal2020}%
  \BibitemOpen
  \bibfield  {author} {\bibinfo {author} {\bibfnamefont {F.}~\bibnamefont {Brauns}}, \bibinfo {author} {\bibfnamefont {J.}~\bibnamefont {Halatek}},\ and\ \bibinfo {author} {\bibfnamefont {E.}~\bibnamefont {Frey}},\ }\href {https://doi.org/10.1103/physrevx.10.041036} {\bibfield  {journal} {\bibinfo  {journal} {Physical Review X}\ }\textbf {\bibinfo {volume} {10}},\ \bibinfo {pages} {041036} (\bibinfo {year} {2020})},\ \Eprint {https://arxiv.org/abs/1812.08684} {1812.08684} \BibitemShut {NoStop}%
\bibitem [{\citenamefont {Brauns}\ \emph {et~al.}(2021{\natexlab{b}})\citenamefont {Brauns}, \citenamefont {Weyer}, \citenamefont {Halatek}, \citenamefont {Yoon},\ and\ \citenamefont {Frey}}]{Brauns.etal2021a}%
  \BibitemOpen
  \bibfield  {author} {\bibinfo {author} {\bibfnamefont {F.}~\bibnamefont {Brauns}}, \bibinfo {author} {\bibfnamefont {H.}~\bibnamefont {Weyer}}, \bibinfo {author} {\bibfnamefont {J.}~\bibnamefont {Halatek}}, \bibinfo {author} {\bibfnamefont {J.}~\bibnamefont {Yoon}},\ and\ \bibinfo {author} {\bibfnamefont {E.}~\bibnamefont {Frey}},\ }\href {https://doi.org/10.1103/physrevlett.126.104101} {\bibfield  {journal} {\bibinfo  {journal} {Physical Review Letters}\ }\textbf {\bibinfo {volume} {126}},\ \bibinfo {pages} {104101} (\bibinfo {year} {2021}{\natexlab{b}})},\ \Eprint {https://arxiv.org/abs/2005.01495} {2005.01495} \BibitemShut {NoStop}%
\bibitem [{\citenamefont {Thalmeier}\ \emph {et~al.}(2016)\citenamefont {Thalmeier}, \citenamefont {Halatek},\ and\ \citenamefont {Frey}}]{Thalmeier.etal2016}%
  \BibitemOpen
  \bibfield  {author} {\bibinfo {author} {\bibfnamefont {D.}~\bibnamefont {Thalmeier}}, \bibinfo {author} {\bibfnamefont {J.}~\bibnamefont {Halatek}},\ and\ \bibinfo {author} {\bibfnamefont {E.}~\bibnamefont {Frey}},\ }\href {https://doi.org/10.1073/pnas.1515191113} {\bibfield  {journal} {\bibinfo  {journal} {Proceedings of the National Academy of Sciences}\ }\textbf {\bibinfo {volume} {113}},\ \bibinfo {pages} {548} (\bibinfo {year} {2016})}\BibitemShut {NoStop}%
\bibitem [{\citenamefont {Ayton}\ \emph {et~al.}(2005)\citenamefont {Ayton}, \citenamefont {McWhirter}, \citenamefont {McMurtry},\ and\ \citenamefont {Voth}}]{Ayton.etal2005}%
  \BibitemOpen
  \bibfield  {author} {\bibinfo {author} {\bibfnamefont {G.~S.}\ \bibnamefont {Ayton}}, \bibinfo {author} {\bibfnamefont {J.~L.}\ \bibnamefont {McWhirter}}, \bibinfo {author} {\bibfnamefont {P.}~\bibnamefont {McMurtry}},\ and\ \bibinfo {author} {\bibfnamefont {G.~A.}\ \bibnamefont {Voth}},\ }\href {https://doi.org/10.1529/biophysj.105.059436} {\bibfield  {journal} {\bibinfo  {journal} {Biophysical Journal}\ }\textbf {\bibinfo {volume} {88}},\ \bibinfo {pages} {3855} (\bibinfo {year} {2005})}\BibitemShut {NoStop}%
\bibitem [{\citenamefont {Goehring}\ \emph {et~al.}(2011)\citenamefont {Goehring}, \citenamefont {Trong}, \citenamefont {Bois}, \citenamefont {Chowdhury}, \citenamefont {Nicola}, \citenamefont {Hyman},\ and\ \citenamefont {Grill}}]{Goehring.etal2011}%
  \BibitemOpen
  \bibfield  {author} {\bibinfo {author} {\bibfnamefont {N.~W.}\ \bibnamefont {Goehring}}, \bibinfo {author} {\bibfnamefont {P.~K.}\ \bibnamefont {Trong}}, \bibinfo {author} {\bibfnamefont {J.~S.}\ \bibnamefont {Bois}}, \bibinfo {author} {\bibfnamefont {D.}~\bibnamefont {Chowdhury}}, \bibinfo {author} {\bibfnamefont {E.~M.}\ \bibnamefont {Nicola}}, \bibinfo {author} {\bibfnamefont {A.~A.}\ \bibnamefont {Hyman}},\ and\ \bibinfo {author} {\bibfnamefont {S.~W.}\ \bibnamefont {Grill}},\ }\href {https://doi.org/10.1126/science.1208619} {\bibfield  {journal} {\bibinfo  {journal} {Science}\ }\textbf {\bibinfo {volume} {334}},\ \bibinfo {pages} {1137} (\bibinfo {year} {2011})}\BibitemShut {NoStop}%
\bibitem [{\citenamefont {Bois}\ \emph {et~al.}(2011)\citenamefont {Bois}, \citenamefont {Jülicher},\ and\ \citenamefont {Grill}}]{Bois.etal2011}%
  \BibitemOpen
  \bibfield  {author} {\bibinfo {author} {\bibfnamefont {J.~S.}\ \bibnamefont {Bois}}, \bibinfo {author} {\bibfnamefont {F.}~\bibnamefont {Jülicher}},\ and\ \bibinfo {author} {\bibfnamefont {S.~W.}\ \bibnamefont {Grill}},\ }\href {https://doi.org/10.1103/physrevlett.106.028103} {\bibfield  {journal} {\bibinfo  {journal} {Physical Review Letters}\ }\textbf {\bibinfo {volume} {106}},\ \bibinfo {pages} {028103} (\bibinfo {year} {2011})}\BibitemShut {NoStop}%
\bibitem [{\citenamefont {Nishikawa}\ \emph {et~al.}(2017)\citenamefont {Nishikawa}, \citenamefont {Naganathan}, \citenamefont {Jülicher},\ and\ \citenamefont {Grill}}]{Nishikawa.etal2017}%
  \BibitemOpen
  \bibfield  {author} {\bibinfo {author} {\bibfnamefont {M.}~\bibnamefont {Nishikawa}}, \bibinfo {author} {\bibfnamefont {S.~R.}\ \bibnamefont {Naganathan}}, \bibinfo {author} {\bibfnamefont {F.}~\bibnamefont {Jülicher}},\ and\ \bibinfo {author} {\bibfnamefont {S.~W.}\ \bibnamefont {Grill}},\ }\href {https://doi.org/10.7554/elife.19595} {\bibfield  {journal} {\bibinfo  {journal} {eLife}\ }\textbf {\bibinfo {volume} {6}},\ \bibinfo {pages} {e19595} (\bibinfo {year} {2017})}\BibitemShut {NoStop}%
\bibitem [{\citenamefont {Klughammer}\ \emph {et~al.}(2018)\citenamefont {Klughammer}, \citenamefont {Bischof}, \citenamefont {Schnellbächer}, \citenamefont {Callegari}, \citenamefont {Lénárt},\ and\ \citenamefont {Schwarz}}]{Klughammer.etal2018}%
  \BibitemOpen
  \bibfield  {author} {\bibinfo {author} {\bibfnamefont {N.}~\bibnamefont {Klughammer}}, \bibinfo {author} {\bibfnamefont {J.}~\bibnamefont {Bischof}}, \bibinfo {author} {\bibfnamefont {N.~D.}\ \bibnamefont {Schnellbächer}}, \bibinfo {author} {\bibfnamefont {A.}~\bibnamefont {Callegari}}, \bibinfo {author} {\bibfnamefont {P.}~\bibnamefont {Lénárt}},\ and\ \bibinfo {author} {\bibfnamefont {U.~S.}\ \bibnamefont {Schwarz}},\ }\href {https://doi.org/10.1371/journal.pcbi.1006588} {\bibfield  {journal} {\bibinfo  {journal} {PLoS computational biology}\ }\textbf {\bibinfo {volume} {14}},\ \bibinfo {pages} {e1006588} (\bibinfo {year} {2018})}\BibitemShut {NoStop}%
\bibitem [{\citenamefont {Illukkumbura}\ \emph {et~al.}(2020)\citenamefont {Illukkumbura}, \citenamefont {Bland},\ and\ \citenamefont {Goehring}}]{Illukkumbura.etal2020}%
  \BibitemOpen
  \bibfield  {author} {\bibinfo {author} {\bibfnamefont {R.}~\bibnamefont {Illukkumbura}}, \bibinfo {author} {\bibfnamefont {T.}~\bibnamefont {Bland}},\ and\ \bibinfo {author} {\bibfnamefont {N.~W.}\ \bibnamefont {Goehring}},\ }\href {https://doi.org/10.1016/j.ceb.2019.10.005} {\bibfield  {journal} {\bibinfo  {journal} {Current Opinion in Cell Biology}\ }\textbf {\bibinfo {volume} {62}},\ \bibinfo {pages} {123} (\bibinfo {year} {2020})}\BibitemShut {NoStop}%
\bibitem [{\citenamefont {Caballero}\ \emph {et~al.}(2023)\citenamefont {Caballero}, \citenamefont {Kruse},\ and\ \citenamefont {Giamarchi}}]{Caballero.etal2023}%
  \BibitemOpen
  \bibfield  {author} {\bibinfo {author} {\bibfnamefont {N.}~\bibnamefont {Caballero}}, \bibinfo {author} {\bibfnamefont {K.}~\bibnamefont {Kruse}},\ and\ \bibinfo {author} {\bibfnamefont {T.}~\bibnamefont {Giamarchi}},\ }\href {https://doi.org/10.1103/physreve.108.l012801} {\bibfield  {journal} {\bibinfo  {journal} {Physical Review E}\ }\textbf {\bibinfo {volume} {108}},\ \bibinfo {pages} {L012801} (\bibinfo {year} {2023})},\ \Eprint {https://arxiv.org/abs/2205.03306} {2205.03306} \BibitemShut {NoStop}%
\bibitem [{\citenamefont {Alimohamadi}\ and\ \citenamefont {Rangamani}(2018)}]{Alimohamadi.Rangamani2018}%
  \BibitemOpen
  \bibfield  {author} {\bibinfo {author} {\bibfnamefont {H.}~\bibnamefont {Alimohamadi}}\ and\ \bibinfo {author} {\bibfnamefont {P.}~\bibnamefont {Rangamani}},\ }\href {https://doi.org/10.3390/biom8040120} {\bibfield  {journal} {\bibinfo  {journal} {Biomolecules}\ }\textbf {\bibinfo {volume} {8}},\ \bibinfo {pages} {120} (\bibinfo {year} {2018})}\BibitemShut {NoStop}%
\bibitem [{\citenamefont {Tozzi}\ \emph {et~al.}(2019)\citenamefont {Tozzi}, \citenamefont {Walani},\ and\ \citenamefont {Arroyo}}]{Tozzi.etal2019}%
  \BibitemOpen
  \bibfield  {author} {\bibinfo {author} {\bibfnamefont {C.}~\bibnamefont {Tozzi}}, \bibinfo {author} {\bibfnamefont {N.}~\bibnamefont {Walani}},\ and\ \bibinfo {author} {\bibfnamefont {M.}~\bibnamefont {Arroyo}},\ }\href {https://doi.org/10.1088/1367-2630/ab3ad6} {\bibfield  {journal} {\bibinfo  {journal} {New Journal of Physics}\ }\textbf {\bibinfo {volume} {21}},\ \bibinfo {pages} {093004} (\bibinfo {year} {2019})}\BibitemShut {NoStop}%
\bibitem [{\citenamefont {Adams}\ and\ \citenamefont {Fournier}(2003)}]{Adams.Fournier2003}%
  \BibitemOpen
  \bibfield  {author} {\bibinfo {author} {\bibfnamefont {R.~A.}\ \bibnamefont {Adams}}\ and\ \bibinfo {author} {\bibfnamefont {J.~J.~F.}\ \bibnamefont {Fournier}},\ }\href@noop {} {\emph {\bibinfo {title} {{Sobolev Spaces}}}},\ \bibinfo {edition} {2nd}\ ed.\ (\bibinfo  {publisher} {Elsevier},\ \bibinfo {year} {2003})\BibitemShut {NoStop}%
\end{thebibliography}%
\end{document}